\documentclass{aa}

\usepackage[utf8]{inputenc}
\usepackage[varg]{txfonts}
\usepackage[breaklinks, colorlinks, citecolor=blue, linkcolor=blue]{hyperref}
\usepackage[usenames, dvipsnames]{color}
\usepackage{amsmath}
\usepackage{graphicx}
\usepackage[english]{babel}
\usepackage{siunitx}
\usepackage{float}
\usepackage{url}
\usepackage{longtable}
\usepackage{fancyhdr}
\usepackage{footmisc}
\usepackage{natbib}
\usepackage[normalem]{ulem}
\usepackage{caption}
\usepackage{multirow}
\usepackage{color}
\usepackage{array} 
\usepackage{ulem}
\usepackage[flushleft]{threeparttable}

\makeatletter
\def\instrefs#1{{\def\scsep{\def\scsep{,}}\@for\w:=#1\do{\scsep\ref{inst:\w}}}}
\renewcommand{\inst}[1]{\unskip$^{\instrefs{#1}}$}

\renewcommand*\aa@pageof{, page \thepage{} of \pageref*{LastPage}} 

\makeatother

\begin{document}

\title{Transmission spectroscopy of MASCARA-1b with ESPRESSO:\\ Challenges of overlapping orbital and Doppler tracks \thanks{Based on guaranteed time observations (GTOs) collected at the European Southern Observatory (ESO) under ESO programme 1102.C-0744 by the ESPRESSO Consortium.}}
\titlerunning{Transmission spectroscopy of MASCARA-1b with ESPRESSO}
\author{N.~Casasayas-Barris\inst{lo}
        \and
        F.~Borsa \inst{INAF_Brera}
        \and
        E.~Palle \inst{iac,ull}
        \and
        R.~Allart \inst{mont,OG}
        \and
        V.~Bourrier \inst{OG}
        \and
        J. I.~Gonz\'alez Hern\'andez \inst{iac,ull}
        \and
        A.~Kesseli \inst{lo}
        \and
        A.~S\'anchez-L\'opez \inst{lo}
        \and       
        M.~R. Zapatero Osorio \inst{calb} 
        \and 
        I. A. G.~Snellen \inst{lo}
        \and
        J.~Orell-Miquel \inst{iac,ull}
        \and
        M.~Stangret \inst{iac,ull}
        \and 
        E.~Esparza-Borges \inst{iac,ull}
        \and 
        C. Lovis \inst{OG}
        \and 
        M.~Hooton \inst{bern}
        \and 
        M.~Lend \inst{OG}
        \and 
        A. M. S.~Smith \inst{gac}
        \and
        F. Pepe \inst{OG}
        \and
        R. Rebolo \inst{iac,ull}
        \and
        S. Cristiani \inst{inaf_Trieste}
        \and        
        N.~C.~Santos \inst{insti_porto,uni_porto}
        \and
        V.~Adibekyan \inst{insti_porto,uni_porto}
        \and
        Y.~Alibert \inst{bern}
        \and
        E.~Cristo \inst{insti_porto,uni_porto}
        \and
        O. D. S.~Demangeon \inst{insti_porto,uni_porto}
        \and
        P.~Figueira\inst{eso,insti_porto}
        \and
        P.~Di~Marcantonio \inst{inaf_Trieste}
        \and
        C.~J.~A.~P.~Martins \inst{insti_porto,centro_porto}
        \and
        G.~Micela \inst{inaf_pal}
        \and
        J.~V.~Seidel \inst{eso}
        \and
        T.~Azevedo Silva \inst{insti_porto,uni_porto}
        \and
        S.~G.~Sousa\inst{insti_porto}
        \and
        A. Sozzetti \inst{inaf_torino}
        \and
        A. Suárez Mascareño \inst{iac,ull}
        \and
        H.~M.~Tabernero \inst{calb}
        }

\institute{
\label{inst:lo}Leiden Observatory, Leiden University, Postbus 9513, 2300 RA Leiden, The Netherlands; \email{barris@strw.leidenuniv.nl}
\and 
\label{inst:INAF_Brera}INAF - Osservatorio Astronomico di Brera, Via Bianchi 46, 23807 Merate, Italy
\and
\label{inst:iac}Instituto de Astrof\'isica de Canarias (IAC), 38205 La Laguna, Tenerife, Spain
\and 
\label{inst:ull}Departamento de Astrof\'isica, Universidad de La Laguna (ULL), 38206, La Laguna, Tenerife, Spain
\and
\label{inst:mont}Department of Physics, and Institute for Research on Exoplanets, Universit\'e de Montr\'eal, Montr\'eal, H3T 1J4, Canada
\and
\label{inst:OG}Observatoire Astronomique de l'Universit\'e de Gen\`eve, Chemin Pegasi 51b, Sauverny, CH-1290, Switzerland
\and
\label{inst:calb}Centro de Astrobiología (CSIC-INTA), Carretera de Ajalvir, km 4. E-28850 Torrejón de Ardoz, Madrid, Spain
\and
\label{inst:bern}Physikalisches Institut, University of Bern, Gesellsschaftstrasse 6, 3012 Bern, Switzerland
\and
\label{inst:gac}Institute of Planetary Research, German Aerospace Center (DLR), Rutherfordstrasse 2, 12489 Berlin, Germany
\and
\label{inst:inaf_Trieste}INAF - Osservatorio Astronomico di Trieste, via G. B. Tiepolo 11, I-34143 Trieste, Italy
\and
\label{inst:insti_porto}Instituto de Astrof\'{\i}sica e Ci\^encias do Espa\c co, Universidade do Porto, CAUP, Rua das Estrelas, 4150-762, Porto, Portugal
\and
\label{inst:uni_porto}Departamento de F\'{\i}sica e Astronomia, Faculdade de Ci\^encias, Universidade do Porto, Rua Campo Alegre, 4169-007, Porto, Portugal
\and
\label{inst:centro_porto}Centro de Astrof\'{\i}sica, Universidade do Porto, Rua das Estrelas, 4150-762 Porto, Portugal
\and
\label{inst:bern}Physics Institute, University of Bern, Sidlerstrasse 5, 3012 Bern, Switzerland
\and
\label{inst:inaf_pal}INAF – Osservatorio Astronomico di Palermo, Piazza del Parlamento 1, 90134 Palermo, Italy
\and
\label{inst:eso}European Southern Observatory (ESO), Alonso de Córdova 3107, Vitacura, Casilla 19001, Santiago de Chile, Chile
\and
\label{inst:inaf_torino}INAF – Osservatorio Astrofisico di Torino, Via Osservatorio 20, 10025 Pino Torinese, Italy}

\date{}

\abstract{
 Atmospheric studies at high spectral resolution have shown the presence of molecules, neutral and ionised metals, and hydrogen in the transmission spectrum of ultra-hot Jupiters, and have started to probe the dynamics of their atmospheres.  We analyse the transmission spectrum of MASCARA-1b, one of the densest ultra-hot Jupiters orbiting a bright (V=8.3) star. We focus on the \ion{Ca}{ii} H\&K, \ion{Na}{i} doublet, \ion{Li}{i}, H$\alpha$, and \ion{K}{i} D1 spectral lines and on the cross-correlated \ion{Fe}{i}, \ion{Fe}{ii}, \ion{Ca}{i}, \ion{Y}{i}, \ion{V}{i}, \ion{V}{ii}, CaH, and TiO lines. For those species that are not present in the stellar spectrum, no detections are reported, but we are able to measure upper limits with an excellent precision ($\sim10$\,ppm for particular species) thanks to the signal-to-noise ratio (S/N) achieved with Echelle SPectrograph for Rocky Exoplanets and Stable Spectroscopic Observations (ESPRESSO) observations. For those species that are present in the stellar spectrum and whose planet-occulted spectral lines induce spurious features in the planetary transmission spectrum, an accurate modelling of the Rossiter-McLaughlin effect (RM) and centre-to-limb variations (CLV) is necessary to recover possible atmospheric signals. In the case of MASCARA-1b, this is difficult due to the overlap between the radial velocities of the stellar surface regions occulted by MASCARA-1b and the orbital track along which the planet atmospheric signal is expected to be found. To try to disentangle a possible planetary signal, we compare our results with models of the RM and CLV effects, and estimate the uncertainties of our models depending on the different system parameters. Unfortunately, more precise measurements of the spin-orbit angle are necessary to better constrain the planet-occulted track and correct for the transit effects in the transmission spectrum with enough precision to be able to detect or discard possible planetary absorptions. Finally, we discuss the possibility that non-detections are related to the low absorption expected for a high surface gravity planet such as MASCARA-1b. Other techniques such as emission spectroscopy may be more useful for exploring their atmospheric composition.
}



\maketitle

\section{Introduction}

During the last few years, ultra-hot Jupiters have become benchmark objects in the field of atmospheric characterisation of exoplanets at high spectral resolution due to their extreme temperatures ($> 2000$\,K; \citealt{Parmentier2018}). Due to their close-in and tidally locked orbits, the temperature in the day-side atmosphere is higher than 2000\,K while the nightside can show temperatures that are significantly lower, leading to a different atmospheric chemistry on both hemispheres of the atmosphere \citep{Arcangeli2018}.

Atmospheric studies of these planets using transmission spectroscopy have shown the presence of neutral and ionised metals such as \ion{Ca}{i}, \ion{Ca}{ii}, \ion{Na}{i}, \ion{Fe}{i}, \ion{Fe}{ii}, \ion{Li}{i}, \ion{K}{i}, \ion{Mg}{i}, and \ion{O}{i} (e.g. \citealt{Hoeijmakers2018, Casasayas2019,Stangret2020,Tabernero2021b,Borsa2021,Seidel2019,Borsa2021Oxy}), as well as hydrogen (e.g. \citealt{YanKELT9,Yan2019}), but no metal hydrides have been detected at high spectral resolution yet (e.g. \citealt{Kesseli2020}). Using observations before and after the secondary
eclipse, Fe, OH, and TiO have been detected in emission from their day-side atmospheres \citep{Yan2020,Pino2020,Nugroho2020FeIe,Cont2021}, which is  direct evidence of thermal inversion layers. Detailed studies with the Echelle SPectrograph for Rocky Exoplanets and Stable Spectroscopic Observations (ESPRESSO; \citealt{Pepe2021}) at the Very Large Telescope (VLT) have revealed asymmetries in the \ion{Fe}{i} atmospheric signals of the ultra-hot Jupiter WASP-76b \citep{Ehrenreich2020}, interpreted by the authors as the combination of time variations produced by atmospheric dynamics and condensation on the night side of the exoplanet. Recently, global circulation modelling explored the origin of this asymmetry and found that the presence of either \ion{Fe}{i} condensation or clouds on the morning side are necessary to explain the asymmetry \citep{Wardenier2021,Savel2021}. This same asymmetry was detected using the High Accuracy Radial velocity Planet Searcher (HARPS) at the La Silla 3.6\,m telescope by \citet{Kesseli2021} and in WASP-121b by \citet{Bourrier2020_HEARTSIII}.

\begin{figure}[]
\centering
\includegraphics[width=0.5\textwidth]{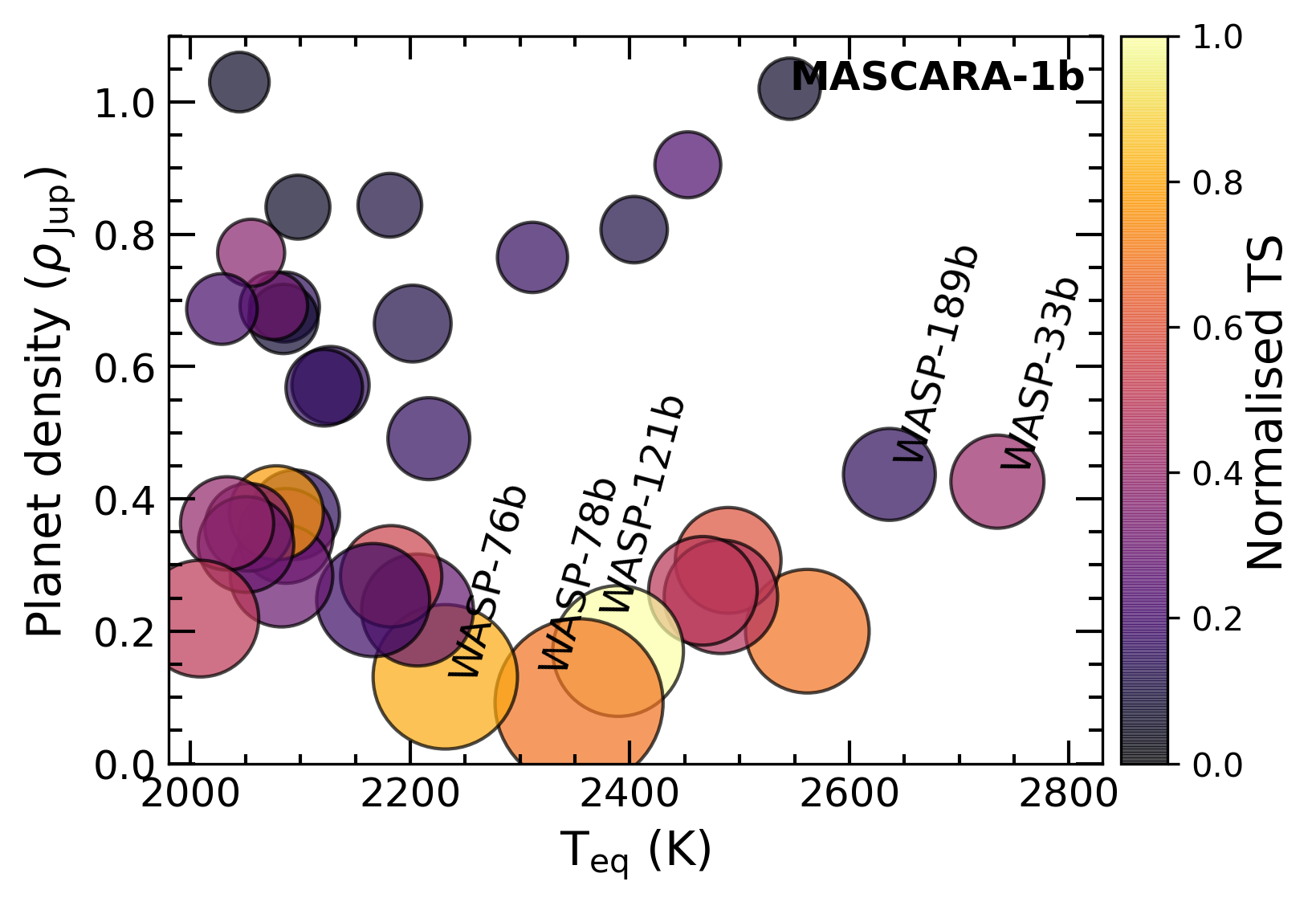}
\caption{Density of the ultra-hot Jupiters with known masses as a function of their equilibrium temperatures ($T_{\rm eq}>2000$\,K). The colour bar shows the ratio of the exoplanet atmosphere annulus surface  and the stellar disc area (TS), considering one scale height ($H$). 
The TS is shown normalised for the sample of exoplanets shown: the TS expands from 0 (lowest transmission) to 1 (highest transmission). The size of the markers is indicative of the $H$ of each exoplanet. Very high-density ultra-hot Jupiters ($> 4\rho_{\rm Jup}$) such as WASP-18b and KELT-9b (with temperature $>3000$\,K) have been discarded for a better visualisation of the sample. The values are calculated with the information extracted from the Transiting Extrasolar Planets catalogue (TEPCat; \citealt{TEPCat}). }
\label{fig:ContextMASC1}
\end{figure}

Here, we explore two transit observations of MASCARA-1b obtained with ESPRESSO to study the atmosphere of this particular ultra-hot Jupiter. MASCARA-1b, discovered by \citet{Talens2017MASC1}, has a day-side temperature of ${\sim3000}$\,K \citep{Bell2021,Hooton2021}, and it is on a misaligned orbit of ${\sim2}$\,days around a bright ($V=8.3$) and fast-rotating ($v\sin i_{\star}=102$\,km\,s$^{-1}$) A8-type star (see more details in Table~\ref{tab:params}). MASCARA-1b is a massive ($3.7~M_{\rm Jup}$) planet with a radius of $1.5~R_{\rm Jup}$, which results in a density very similar to that of Jupiter. In the context of the ultra-hot Jupiters known to date, MASCARA-1b is one of the densest exoplanets of this category (see Figure~\ref{fig:ContextMASC1}). Recent CHEOPS (CHaracterising ExOPlanets Satellite) observations \citep{Hooton2021} have shown asymmetries in the transit resulting from the gravity darkening induced by the stellar rotation. \citet{Stangret2021} studied the atmosphere of MASCARA-1b using four HARPS-N transit observations, reporting non-detections in the transmission spectrum of the planet and showing the overlap of both planet and Doppler shadow radial velocities. Due to its density, MASCARA-1b is expected to show faint atmospheric signatures in the transmission spectrum. However, the brightness of its host star and the strong stellar residuals make MASCARA-1b the perfect exoplanet to assess the current precision in the data analysis and the accuracy in the modelling.

\begin{table}[]
\small
\centering
\caption{Physical and orbital parameters of the MASCARA-1b system.}
\begin{tabular}{lrl}
\hline \hline
\\[-1em]
 Parameter  & Value & Reference$^a$\\ \hline
 \\[-1em]
  \multicolumn{3}{c}{\dotfill\it Stellar parameters \dotfill}\\\noalign{\smallskip}
   \\[-1em]
\quad  $T_{\rm eff}$ [K] & $7554\pm 150$  & T2017 \\
  \\[-1em]
\quad  $\log g$ [cgs]& $4$ & T2017\\
  \\[-1em]
\quad  [Fe/H] [dex] & $0$ & T2017\\
  \\[-1em]
\quad $R_{\star}$ [$\rm{R_{\odot}}$]& $2.082^{+0.022}_{-0.024}$ & H2021\\
  \\[-1em]
 \quad  $v\sin i_{\star}$ [km\,s$^{-1}$]& $101.7^{+3.5}_{-4.2}$ & H2021\\
  \\[-1em]
  \multicolumn{3}{c}{\dotfill\it Planet parameters \dotfill}\\\noalign{\smallskip}
   \\[-1em]
 \quad $M_{\rm p}$ [$\rm{M_{Jup}}$]&  $3.7\pm0.9$ & T2017\\
  \\[-1em]
 \quad $R_{\rm p}$ [$\rm{R_{Jup}}$]& $1.597^{+0.018}_{-0.019}$ & H2021\\
 \\[-1em]
\quad $K_{\rm p}$ [km\,s$^{-1}$]& $204.2\pm0.3$ & This work\\
  \\[-1em]
  \multicolumn{3}{c}{\dotfill\it Transit parameters \dotfill}\\\noalign{\smallskip}
   \\[-1em]
 \quad $T_{\rm 0}$ [BJD$_{\rm TDB}$] & $2458833.488151^{+0.000091}_{-0.000092}$ & H2021\\
  \\[-1em]
\\[-1em]
 \quad $P$ [d] & $2.14877381^{+0.00000088}_{-0.00000087}$ & H2021\\
  \\[-1em]
 \quad $T_{14}$ [hr] & $4.226^{+0.010}_{-0.011}$ & H2021\\
 \\[-1em]
  \multicolumn{3}{c}{\dotfill\it System parameters \dotfill}\\\noalign{\smallskip}
   \\[-1em]
  \quad $a/R_{\star}$& $4.1676^{+0.0047}_{-0.0051}$ & H2021 \\
 \\[-1em]
   \quad $a$ [au]& $0.040352^{+0.000046}_{-0.000049}$ & H2021 \\
 \\[-1em]
 \quad $R_{\rm p}/R_{\star}$& $0.07884^{+0.00022}_{-0.00021}$ & H2021\\
 \\[-1em]
 \quad $i_{\rm p}$ [deg]& $88.45\pm0.17$ & H2021\\
 \\[-1em]
 \quad $e$& $0.00034^{+0.00034}_{-0.00033}$ & H2021\\
\\[-1em]
\quad $K_{\star}$ [m\,s$^{-1}$]& $400\pm100$ & T2017\\
  \\[-1em]
 \quad $\lambda$ [deg]& $69.2^{+3.1}_{-3.4}$ & H2021\\
   \\[-1em]
\quad $v_{\rm sys}$ [km\,s$^{-1}$]& $9.3\pm2.3$ & This work\\
\\[-1em]
\lasthline
\end{tabular}
\tablefoot{\tablefoottext{a} T2017 corresponds to \citet{Talens2017MASC1} and H2021 to \citet{Hooton2021}.}
\label{tab:params}
\end{table}

This paper is organised as follows. In Section~\ref{sec:observations} we describe the observations and data reduction. In Section~\ref{sec:sysv} we explore the systemic velocity of MASCARA-1. In Section~\ref{sec:analysis} we describe the methodology used to extract and explore the transmission spectrum of the exoplanet. In Section~\ref{sec:results} the results from the atmospheric analysis are shown and discussed. In Section~\ref{sec:conclusions} we summarise the results and conclusions.

\section{Observations and data reduction} \label{sec:observations}

Two transits of MASCARA-1b were observed on the nights of July 13, 2019,  and August 10, 2019, using the ESPRESSO spectrograph as part of the guaranteed time observation (GTO) under programme 1102.C-0744. ESPRESSO is a fibre-fed spectrograph located at the Very Large Telescope (VLT) that covers the optical range between 3800 and 7880$~{\rm \AA}$. The observations were performed at the UT3 Melipal telescope and using the HR21\footnote{HR21 considers 1-arcsec fibre and a binning of a factor of 2 along the spatial direction.} observing mode, achieving a resolving power of $\Re\sim140~000$ \citep{Pepe2021}.

\begin{figure}[]
\centering
\includegraphics[width=0.45\textwidth]{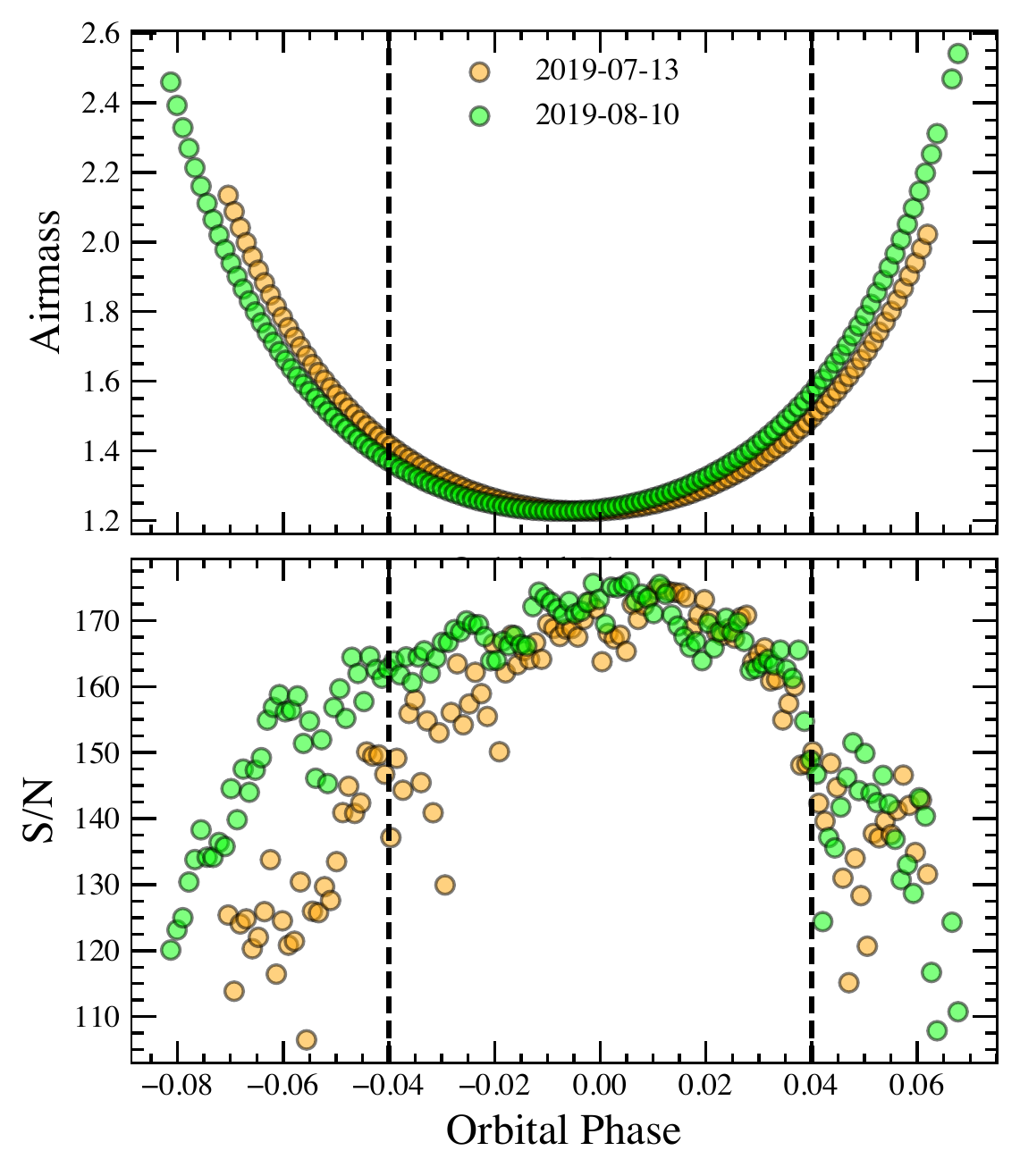}
\caption{Evolution of the airmass (top panel) and a S/N around 590\,nm (bottom panel) during the two transit observations with ESPRESSO. The two vertical dashed lines indicate the first and last contacts of the transit.}
\label{fig:SNR}
\end{figure}

During the two nights of observation, the star was monitored taking consecutive exposures of 150\,s before, during, and after the transit of the exoplanet. Fibre A was used to observe the target and fibre B to simultaneously monitor the sky. In total, 117 and 130 exposures were obtained for the first (July 13, 2019) and second (August 10, 2019) night. This was ${\sim7}$ and ${\sim7.5}$\,hours of observation, respectively. The average S/N obtained for the first night in the 590\,nm order was  151, while for the second night it was 157. The observations are summarised in Table~\ref{tab:obs}, and the airmass and S/N evolution during the nights of observation are presented in Figure~\ref{fig:SNR}.

The observations were reduced using the data reduction software (DRS) pipeline 2.2.8. Due to the operational limit in the atmospheric dispersion corrector, the exposures with an airmass higher than 2.2 are excluded from the analysis presented here \citep{Allart2020}. This affects nine exposures of the second night (August 10, 2019). After excluding these exposures, the averaged S/N for this night is 159.

\begin{table*}[]
\centering
\caption{Observing log of MASCARA-1b transit observations.}
\begin{tabular}{ccccccccc}
\hline\hline
Night & Date of   & Telescope     & Start & End &  Airmass$^a$ &$T_\mathrm{exp}$ & $N_\mathrm{obs}$ & S/N$^b$ \\
         & observation & & [UT] & [UT] & change & [s] &         &    \\ \hline
\\[-1em]
1 & 2019-07-13 & VLT-UT3 & 02:56 & 09:46 & 2.13-1.23-2.02 & 150 & 117 &  106-175\\
\\[-1em]
2 & 2019-08-10 & VLT-UT3 &  00:47 &  08:28 & 2.46-1.23-2.54 & 150 & 130 & 108-176\\ 
\\[-1em]
\hline
\end{tabular}\\
\tablefoot{\tablefoottext{a}{Airmass change during the observation.} \tablefoottext{b}{Minimum and maximum S/N for each night, calculated in the Na I echelle order.}}
\label{tab:obs}
\end{table*}

\section{The systemic velocity of MASCARA-1} \label{sec:sysv}

\begin{figure}[]
\centering
\includegraphics[width=0.5\textwidth]{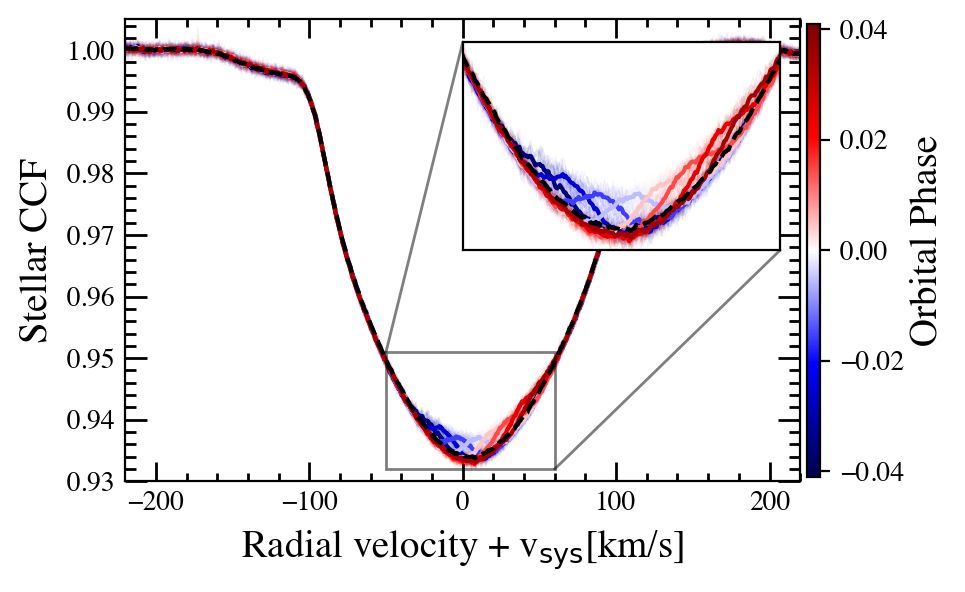}
\caption{Observed stellar CCFs of MASCARA-1 obtained by cross-correlating the full ESPRESSO spectrum with an A0 mask during the transit without correcting the systemic velocity. Both nights are shown together, ordered according to the orbital phase of the planet, which is indicated in different colours. The thick coloured lines show the CCFs binned by $0.01$ in orbital phase, and the averaged out-of-transit CCF is shown as a dashed black line. The CCFs have been normalised for a better visualisation. The inset shows a zoomed-in view of the core of the CCFs, where the Doppler shadow of MASCARA-1b can be clearly observed.}
\label{fig:CCFs}
\end{figure}

MASCARA-1 is a fast-rotating A-type star whose spectrum only shows a small number of broad spectral lines. The systemic velocity ($v_{\rm sys}$) of MASCARA-1 was measured by \citet{Talens2017MASC1}, finding different values for different instruments: $v_{\rm sys} = 11.20 \pm 0.08$\,km\,s$^{-1}$ (HERMES) and $v_{\rm sys} = 8.52 \pm 0.02$\,km\,s$^{-1}$ (SONG). Here, we estimate $v_{\rm sys}$ by cross-correlating the combined out-of-transit ESPRESSO observations with a synthetic spectrum of MASCARA-1 computed using the MARCS stellar models \citep{MARCS2008A&A...486..951G}, the spectroscopy made easy tool \citep{SMEEvolution2017}, and the stellar parameters from Table~\ref{tab:params}. We select the wavelength region 4970--5830~$\rm \AA$, which includes a relatively large number of spectral lines and excludes very broad lines such as \ion{H}{i} Balmer lines and \ion{Na}{i}. Before applying the cross-correlation, both model and observations were normalised with a polynomial of order 5 and re-binned with a pixel size of $0.51$\,km\,s$^{-1}$. The cross-correlation function (CCF) was constructed in the range between 5050 and 5650~$\rm \AA$ and the centre of this CCF was then determined with a parabolic fit, obtaining $v_{\rm sys} = 9.3\pm2.3$\,km\,s$^{-1}$, which is consistent with previous results. However, if we look at the stellar CCFs obtained with the ESPRESSO DRS using an A0 mask, we find an average out-of-transit velocity of $\sim3.8$\,km\,s$^{-1}$ (see the extracted CCFs in Fig.~\ref{fig:CCFs}). 

Absolute radial-velocity measurements of hot fast-rotating stars are challenging, as only a small number of extremely blended spectral lines are available. This might produce the differences observed when using different methodologies as late type stars have shown consistent radial-velocity measurements (see \citealt{Faria2022}). In the atmospheric analysis presented in Sec.~\ref{sec:analysis}, we do not correct for this radial-velocity offset. Instead, in the results, we explore the wavelength region where the possible atmospheric signal is expected. 

\section{Methods} \label{sec:analysis}

In the analysis presented here, we used the one-dimensional spectra (S1D sky subtracted products), which have been corrected with {\tt Molecfit} as presented by \citet{Allart2017} and \citet{Tabernero2021b}. The telluric correction can be observed in the top panel of Fig.~\ref{fig:CC_mods}. As expected, we achieved the noise level correction in most of the ESPRESSO wavelength coverage except for those telluric regions containing saturated absorption lines, such as the O$_2$ lines. We searched for different atomic and molecular species in the atmosphere of MASCARA-1b using the cross-correlation technique and the analysis of particular single lines.

\subsection{Survey of atomic and molecular species using the cross-correlation technique} \label{sec:cc_analy}

\subsubsection{Atmospheric models of MASCARA-1b} \label{sec:atm_models}

The atmospheric model templates of MASCARA-1b used in this study are generated using the {\tt petitRADTRANS} code \citep{petitRADTRANS2019}. Following previous studies of ultra-hot Jupiters (e.g. \citealt{Hoeijmakers2019, Stangret2020, Borsa2021}), for this calculation, we assumed the planet parameters presented in Table~\ref{tab:params}, solar abundance, a continuum pressure level of 1\,mbar, and an isothermal atmosphere at 3000\,K, which is slightly higher than the equilibrium temperature of MASCARA-1b. The atmospheric models were then convolved at the ESPRESSO resolving power. 

We explored the presence of \ion{Fe}{i}, \ion{Fe}{ii}, \ion{Ca}{i}, \ion{Y}{i}, \ion{V}{i}, \ion{V}{ii}, CaH, and TiO in the atmosphere of MASCARA-1b using the cross-correlation technique. For TiO we used the TOTO line list \citep{TiOTOTO2019}, for CaH we used the line list by \citet{CaHList2016}, and for the other species we used the line lists provided in {\tt petitRADTRANS}. The accuracy of the TiO and CaH line lists has been tested in an M dwarf. Recently, \citet{Prinoth2021} robustly detected TiO in WASP-189b using this same line list. For CaH, we discarded the lines below 6000\,\AA, which show less accuracy. Together with VO, these two species are the dominant sources of opacity from molecules in the optical spectra of M dwarfs of similar temperature. We did not explore the presence of VO here, as several studies have shown that the current line lists are not accurate enough (e.g. \citealt{Merritt2020}). The presence of metal hydrides has been evidenced at low spectral resolution \citep{Skaf2020,Braam2021}, but no detections have been confirmed at high resolution \citep{Kesseli2020}. The atmospheric models for the different species are shown in Fig. \ref{fig:CC_mods}.

\subsubsection{Cross-correlation analysis} \label{sec:cc}

Before cross-correlating the data with the different atmospheric models, we applied several steps. First, we moved all spectra to the stellar rest frame considering the stellar radial-velocity semi-amplitude ($K_{\star}$) presented in Table~\ref{tab:params} and using the {\tt SinRadVel} routine from PyAstronomy \citep{PyAstronomy2019ascl.soft06010C}. This value was not well constrained, but the correction was minor with a change of $0.3$\,km\,s$^{-1}$ between the first and the last spectrum of the night. Then, we normalised the spectra to the same continuum level following the methodology described by \citet{Merritt2020}. This methodology allowed us to correct variations in the continuum (e.g. the wiggles observed in ESPRESSO data by \citealt{Tabernero2021b, Borsa2021} or \citealt{Casasayas2021}). In summary, each spectrum was divided by the median spectrum of the time series. The resulting division was smoothed with a median filter using a width of 15 pixels and a Gaussian filter with a standard deviation of 50 pixels. We obtained the variations in the continuum of each spectrum with respect to the reference spectrum, which are similar to a blaze function correction. Then, each original spectrum was divided by the resulting filter, that is, the variations of the continuum.  

\begin{figure*}[]
\centering
\includegraphics[width=1\textwidth]{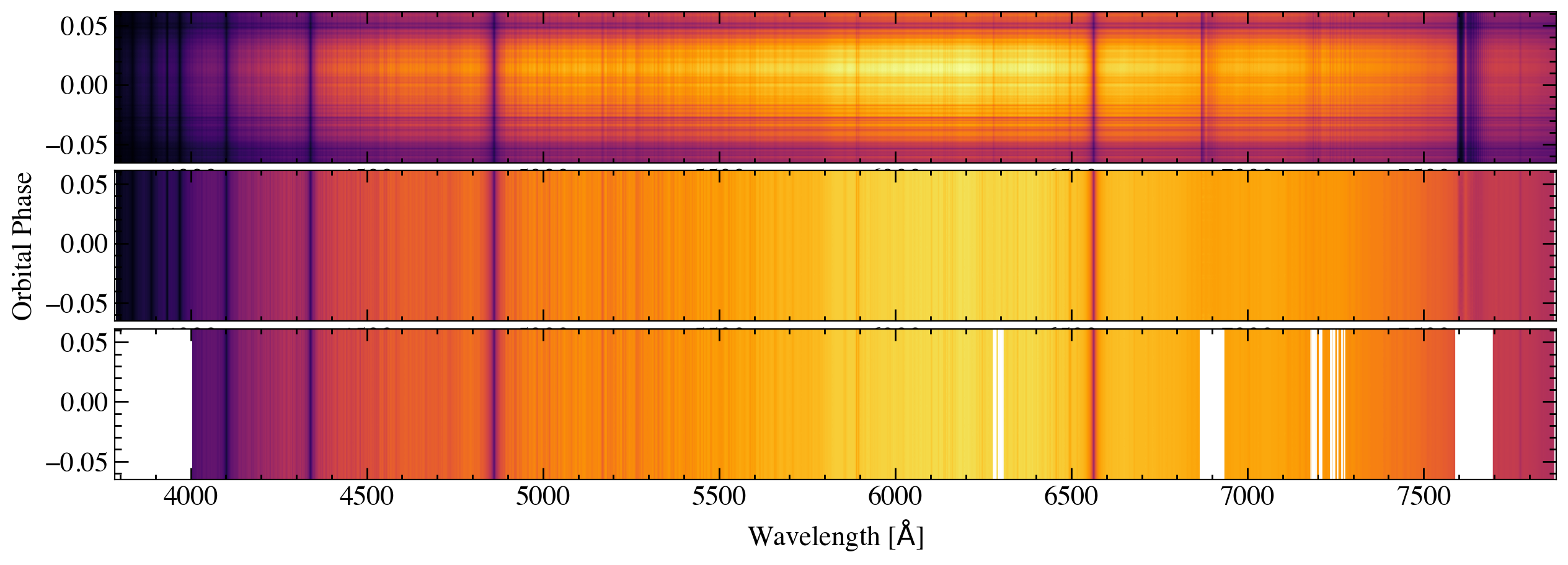}
\caption{Steps applied to the observations before the cross-correlation analysis. As an example, we show the data of the first night. {\em Top panel:} Original time series. {\em Middle panel:} Time series obtained after correcting the telluric contamination and normalising all spectra to the same continuum level as presented in \citet{Merritt2020}. {\em Bottom panel:} Time series of spectra after masking the columns of pixels affected by telluric residuals and the region below 4000\,{\AA} due to a lower S/N.}
\label{fig:Method}
\end{figure*}

After that, we cleaned the observations using a 3$\sigma$ clipping in the time dimension (see \citealt{Hoeijmakers2020}, for example). We computed the standard deviation in the time domain and then rejected those pixels that were three times above this value. We additionally masked individual columns of pixels that were affected by telluric residuals and discarded regions with strong telluric bands (6868--6930\,{\AA}, 7592--7692\,{\AA}) and in the range below 4000\,{\AA} due to a lower S/N. In total, we masked $\sim 13$\% of the pixels in both nights. Most of the masked columns were located in strong telluric contaminated regions. These steps are illustrated in Fig.~\ref{fig:Method}. 

Once the pre-analysis steps were applied we cross-correlated the time series with the atmospheric models presented in Sec.~\ref{sec:atm_models}. We used the CCF presented in \citet{Hoeijmakers2020}:
\begin{equation}
    c(\nu,t) = \sum_{i=0}^{N} x_i(t)T_i(\nu),
\label{eq:ccf}
\end{equation}
where $c(\nu,t)$ is the cross-correlation for a radial-velocity value $\nu$ at the time $t$. $x_i(t)$ is the spectrum from the bottom panel of Fig.~\ref{fig:Method} observed at a given time and $T_i(\nu)$ the model at each velocity. The cross-correlation was applied over a range of $\pm350$\,km\,s$^{-1}$ in steps of 0.5\,km\,s$^{-1}$, following the pixel scale of ESPRESSO without resampling. We normalised the model following $\sum_{i=0}^{N} T_i(\nu) = 1$, so that the pixels ($i$) in the transmission spectrum were weighted by their absorption. It can be observed that Eq.~\ref{eq:ccf} provides the cross-variance, as the normalisation factor of the CCF is missing (see, for example, \citealt{BrogiLine2019}). In this case, if both model and data have the same units, Eq.~\ref{eq:ccf} provides the average line depth of the cross-correlated lines.

The individual CCFs were then divided by the out-of-transit cross-correlation and moved to the planet rest frame using $K_p = 204.2\pm0.3$\,km\,s$^{-1}$ (from $K_p= 2\pi a \sin(i_p)/P$ and the parameters from Table~\ref{tab:params}), and the resulting in-transit cross-correlation values were then co-added. We additionally calculated the $K_p$-velocity maps, shifting the in-transit cross-correlation results for different $K_p$ values in the range $K_p\pm500$\,km\,s$^{-1}$ in steps of 0.5\,km\,s$^{-1}$. The S/N of the signals is estimated relative to the standard deviation calculated in the ranges $-300$--$-200$\,km\,s$^{-1}$, and $+200$--$+300$\,km\,s$^{-1}$, far from the Rosstier-McLaughlin (RM) feature and a possible atmospheric signal.

\subsection{Transmission spectrum around single lines} \label{sec:lines}

We explored the transmission spectrum of MASCARA-1b around the \ion{Ca}{ii} H\&K, \ion{Na}{i} doublet, H$\alpha$, \ion{Li}{i}, and \ion{K}{i} D1 lines, following the methodology presented in \citet{Wytt2015} and \citet{Casasayas2018}. As detailed in Sec.~\ref{sec:cc}, the spectra were first shifted to the stellar rest frame and normalised. We then combined all out-of-transit observations to build a high S/N stellar spectrum and compute the ratio of each individual spectrum by this master stellar spectrum to remove the stellar contribution. The resulting in-transit data were shifted to the planet rest frame and combined to obtain the transmission spectrum of the exoplanet.

In the \ion{Na}{i} and \ion{Ca}{ii} H\&K stellar lines, we observed interstellar medium (ISM) absorption. The ISM is expected to be static during the observations in the frame of the Solar System barycentre. For this reason, we got rid of this contamination when computing the ratio of each individual spectrum by the master stellar spectrum, as the $K_{\star}$ contribution produced a small radial-velocity change of $\sim0.3$\,km\,s$^{-1}$, which was not significant in our observations, especially taking into account the broad spectral lines of MASCARA-1.

\subsection{RM and CLV effects modelling} \label{sec:RMcorr}

Due to MASCARA-1's system architecture and the fast rotation of the stellar host, we observed a strong RM effect (or Doppler shadow) on the stellar lines (see Fig.~\ref{fig:CCFs}) that overlapped with the position where a possible planetary absorption was expected due to their similar radial-velocities. Following the methodology presented by \citet{Yan2017A&A...603A..73Y} and \citet{YanKELT9}, we simulated the observations of MASCARA-1 including the centre-to-limb variation (CLV) and the RM effects. This methodology has been used in several recent atmospheric studies using high spectral resolution observations such as \citet{Chen2020}, \citet{Borsa2021}, and \citet{Casasayas2021}, among others. Here, we used the ATLAS9 \citep{ATLAS92003} stellar models over the whole ESPRESSO wavelength coverage, assuming solid-body rotation, solar abundances, and the values presented in Table~\ref{tab:params} from \citet{Hooton2021}. 
The ultra-high precision of CHEOPS photometry \citep{Benz2021} revealed the signature of oblateness and gravity-darkening in the transit light curve of MASCARA-1b \citep{Hooton2021}, which we ignore here. In principle, this effect could modify the intensity of the RM effect during the transit \citep{Barnes2009}. 

Several studies have shown the possibility of correcting for the RM effect and CLV by fitting a combination of Gaussian profiles (e.g. \citealt{Hoeijmakers2020}). Here, we prefered a more accurate modelling, so the effect of the lines surrounding the species of interest were also estimated. Once the stellar spectra at different orbital phases of the planet were simulated, we followed the exact same process as for the observations (see Sects.~\ref{sec:cc} and~\ref{sec:lines}) to determine the impact of these effects on the results. 

\section{Results and discussion} \label{sec:results}

\subsection{Atmospheric characterisation of MASCARA-1b} \label{sec:res}

\begin{figure*}[]
\centering
\includegraphics[width=1\textwidth]{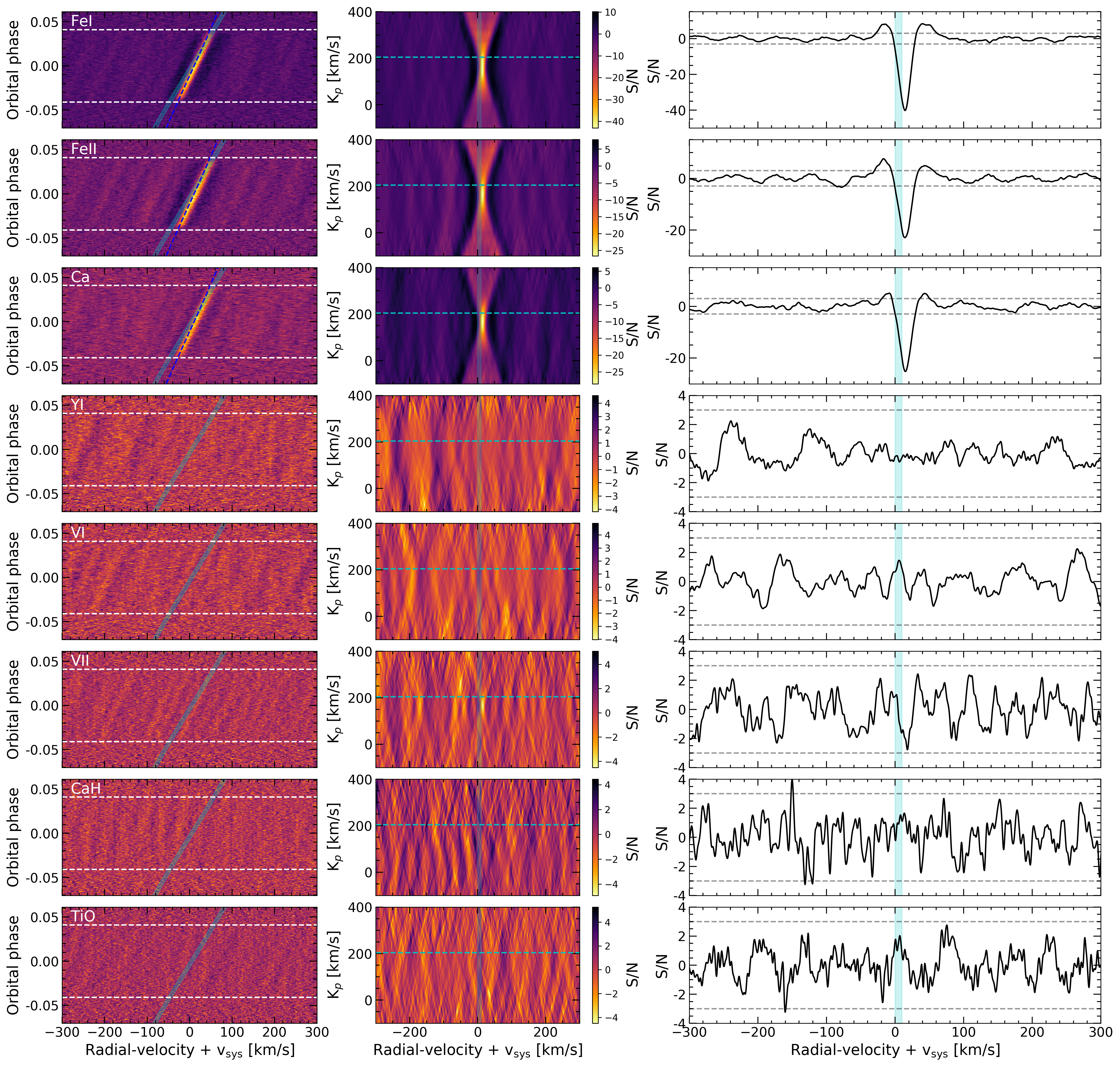}
\caption{MASCARA-1b cross-correlation results obtained with atmospheric models of different species computed with {\tt petitRADTRANS}. Each row shows the results for one species. {\em Left column}: Cross-correlation maps. The first and last contacts of the transit are shown in horizontal dashed white lines. The light cyan region shows the planet radial-velocity paths considering the different possible $v_{\rm sys}$ values for this star ($\sim 3 - 11$\,km\,s$^{-1}$) based on literature measurements and the calculations performed here and the dashed blue line indicates the radial-velocity path of the Doppler shadow. {\em Middle column}: $K_p-v_{\rm sys}$ map. The horizontal dashed line indicates the expected $K_p$ at $204.2$\,km\,s$^{-1}$. The colour bar shows the S/N. {\em Right column}: Combined in-transit CCFs after correcting the planet radial velocities. The horizontal grey dashed lines indicate the $\pm3\sigma$ level. The radial velocities (horizontal axis) of these results do not consider the systemic velocity correction. The strong features observed in the \ion{Fe}{i}, \ion{Fe}{ii}, and \ion{Ca}{i} panels correspond to the Doppler shadow.}
\label{fig:CC_res}
\end{figure*}

\begin{figure*}[]
\centering
\includegraphics[width=0.98\textwidth]{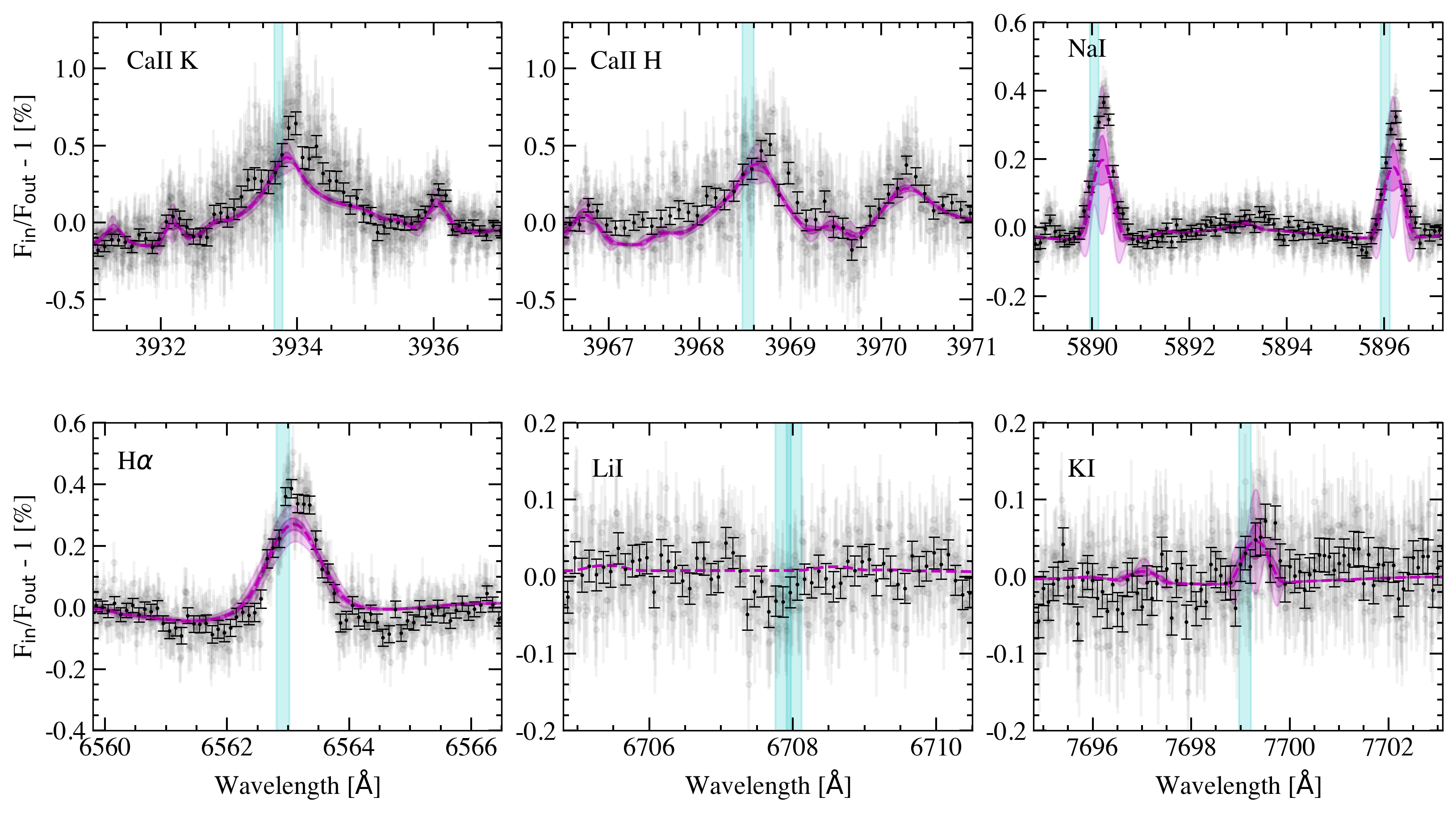}
\caption{Transmission spectrum of MASCARA-1b around single lines, after combining the two nights of observation. In grey  we show the original sampling of the data and in black the data binned by $0.1$\,{\AA}. The width of the vertical light blue region indicates the uncertainty of $v_{\rm sys}$, that is, the position where the signals from the exoplanet atmosphere are expected (we note that the \ion{Li}{i} is a doublet). The estimated models describing the RM and CLV effects on the transmission spectrum and their $1\sigma$ and $3\sigma$ uncertainties are shown in violet.}
\label{fig:TS_res}
\end{figure*}

Using the cross-correlation technique, we searched for \ion{Fe}{i}, \ion{Fe}{ii}, \ion{Ca}{i}, \ion{Y}{i}, \ion{V}{i}, \ion{V}{ii}, CaH, and TiO in ESPRESSO observations of MASCARA-1b. The results are summarised in Fig.~\ref{fig:CC_res}. Those species present in the stellar spectrum (\ion{Fe}{i}, \ion{Fe}{ii}, and \ion{Ca}{i}) are clearly affected by the RM effect, which overlaps with the radial-velocities of the exoplanet during the transit. This makes it extremely challenging to determine if a signal from the exoplanet atmosphere is present or not in the observations (see discussion in Sec.~\ref{sec:RM_disc}). On the other hand, to avoid the overlap with the RM effect, we focussed on species that are not present in the stellar spectrum (\ion{Y}{i}, \ion{V}{i}, \ion{V}{ii}, CaH, and TiO), but none of them are detected in our analysis. Similarly, when focussing on the analysis of single lines, all explored species, except Li, are affected by the presence of the Doppler shadow (see Fig.~\ref{fig:TS_res}). A faint absorption feature of $0.05~\%$ is noticed close to the \ion{Li}{i} ($\sim3\sigma$) lines position. However, we do not consider this feature to be significant enough ($>3\sigma$) to be claimed as a detection.

For those species not present in the stellar spectrum, we measure upper limits of the non-detections. First of all, following \citet{Hoeijmakers2020} we inject the atmospheric models to the observations considering the broadening of the lines produced by the tidally locked rotation of the planet. For this, we convolve MASCARA-1b atmospheric models with a broadening kernel of $v_{\rm rot}=3.9$\,km\,s$^{-1}$ using the rotation broadening routine {\tt fastRotBroad} from Pyastronomy \citep{PyAstronomy2019ascl.soft06010C}. This is only an approximation, as the rotation of the atmospheric annulus would lead to double peaked lines \citep{Brogi2016}, which is not considered here. Similarly, we also take into account the exoplanet movement in one exposure of 150\,s, which corresponds to $v_{\rm exp}=1.0$\,km\,s$^{-1}$, considering $K_p = 204.2$\,km\,s$^{-1}$ calculated with the parameters presented in Table~\ref{tab:params}. We then weight the models along the transit using the photometric light curve derived with {\tt PyTransit} \citep{Parviainen2015Pytransit} and the {\tt LDTk} \citep{Parviainen2015LDTK} to estimate the limb-darkening coefficients of MASCARA-1 in the ESPRESSO wavelength range ($q_1 =0.67\pm0.01$ and $q_2 =0.36\pm0.01$). The injection of these models was performed before masking and normalising the time series.

Once the models have been injected, in order to decrease the model dependence of the upper limit measurements, we cross-correlate the data with a binary mask as presented in \citet{Allart2020}. We analysed the data by increasing the number of lines from the atmospheric models (see Fig.~\ref{fig:CC_mods}) that are included in the binary mask, from the strongest to the faintest. This helps us explore for which number of lines the recovered (injected) signal is more significant and select that number to calculate the upper limits. This significance is obtained by fitting a Gaussian profile to the cross-correlation results. For this iterative calculation, the models are injected at $+135$\,km\,s$^{-1}$ (to avoid a possible atmospheric signal from the planet) and the noise level is calculated in the region $|200-300|$\,km\,s$^{-1}$. The evolution of the S/N of the recovered signal with the number of lines included in the binary mask is shown in Fig.~\ref{fig:Nlines}. For the selected number of lines, we calculate the upper limits following two different approaches. First, we compute the 1-pixel dispersion of the result in the region where the atmospheric feature is expected, between $\pm100$\,km\,s$^{-1}$, and compute the $3\sigma$ upper limit from this value. On the other hand, we use the full width at half maximum (FWHM) of the recovered signal and integrate the 1-pixel dispersion over the correspondent number of pixels to estimate the $3\sigma$ line precision, assuming that the particular model is a good description of the atmosphere. More details on these calculations can be found in \citet{Allart2017} and \citet{Allart2020}.

The resulting upper limits are summarised in Table~\ref{tab:upper_lim1}. For the particular case of \ion{V}{ii}, which contain only $\sim200$ lines in the ESPRESSO wavelength range, the injection could not be significantly recovered. During the upper limits calculation we observed that the effect of the Doppler shadow from surrounding lines were intercepted by the mask in the cross-correlation analysis, affecting the noise level estimation and, consequently, the upper limits. For this reason, we attempted to correct the RM effect of these lines, allowing us to estimate the noise level more accurately (see Fig.~\ref{fig:RM_corr1} and Fig.~\ref{fig:RM_corr2}). For \ion{V}{i} and \ion{V}{ii}, for example, the standard deviation of the continuum (calculated between $|100-300|$\,km\,s$^{-1}$) is 15 and 48\,pmm before, and 12 and 41\,ppm after the correction, respectively. This means that the noise is reduced by $20$\% and $15$\% after the correction. Although the RM modelling might not be completely accurate (see Sec.~\ref{sec:RM_disc}), this correction helps us to reduce the variations of the continuum introduced by the RM of other species and measure the noise level more accurately. Our upper limits show that MASCARA-1b ESPRESSO observations reach a precision of $\sim10$\,ppm for some species.

\begin{table*}[]
\centering
\caption{3$\sigma$ upper limits of the non-detections obtained with the cross-correlation technique.}
\begin{tabular}{lccccccc}
\hline\hline
Species  & Number   & 1-pixel upper    & 1-pixel upper      & Line width & Line upper  & Line upper & S/N\\
 &   of lines & limit [ppm] & limit [H$^a$] & [km\,s$^{-1}$] & limit [ppm] &   limit [H]  & rec.$^b$ \\ \hline
\\[-1em]




\ion{Y}{i} & 870 & 40 | 36$^c$ & 1.4 | 1.3 & 6.1 & 12 | 10 & 0.4 | 0.4 & 3.8 | 4.5 \\

\ion{V}{i} & 1100 & 40 | 35  & 1.4 | 1.2 & 6.4 & 11 | 10 & 0.4 | 0.3 & 4.2 | 5.5 \\
 
\ion{V}{ii} & 100 & 147 | 134 & 5.2 | 4.8 & 5.9$^d$ & 43 | 39 & 1.5 | 1.4 & < 1.0 \\

CaH & 1500 & 26 | 26 & 0.9 | 0.9 & 3.8 & 9 | 9 & 0.3 | 0.3 & 5.6 | 5.8\\

TiO & 6000 & 16 | 14 & 0.6 | 0.5 & 2.5 & 7 | 6 & 0.2 | 0.2 & 3.9 | 4.6\\
  
\hline
\end{tabular}\\
\tablefoot{\tablefoottext{a}{Number of scale heights, considering that $H=258$\,km, which produces an absorption depth of 28\,ppm.} \tablefoottext{b}{S/N at which the injected atmospheric models are recovered.} \tablefoottext{c}{The first value is obtained before correcting for the RM and CLV effects as described in Sec.~\ref{sec:res}, while the second value is obtained after this correction.} \tablefoottext{d}{For \ion{V}{ii,} the signal is not recovered, so the line width is measured after computing the difference between the results obtained with and without the injection. }}
\label{tab:upper_lim1}
\end{table*}

\subsection{Impact of the Doppler shadow on the transmission spectrum} \label{sec:RM_disc} 

MASCARA-1 is a fast-rotating ($v\sin i_{\star} = 102$\,km\,s$^{-1}$) A-type star for which inconsistent systemic velocity values are obtained when using different methodologies, ranging between ${\sim3}$ and ${\sim11}$\,km\,s$^{-1}$. Additionally, due to the system's architecture, the Doppler shadow and the planet radial velocity during the transit overlap. This complicates the study of MASCARA-1b's atmosphere using transmission spectroscopy at high spectral resolution for those species also present in the stellar spectrum \citep{Casasayas2021}.

\begin{figure*}[]
\centering
\includegraphics[width=0.98\textwidth]{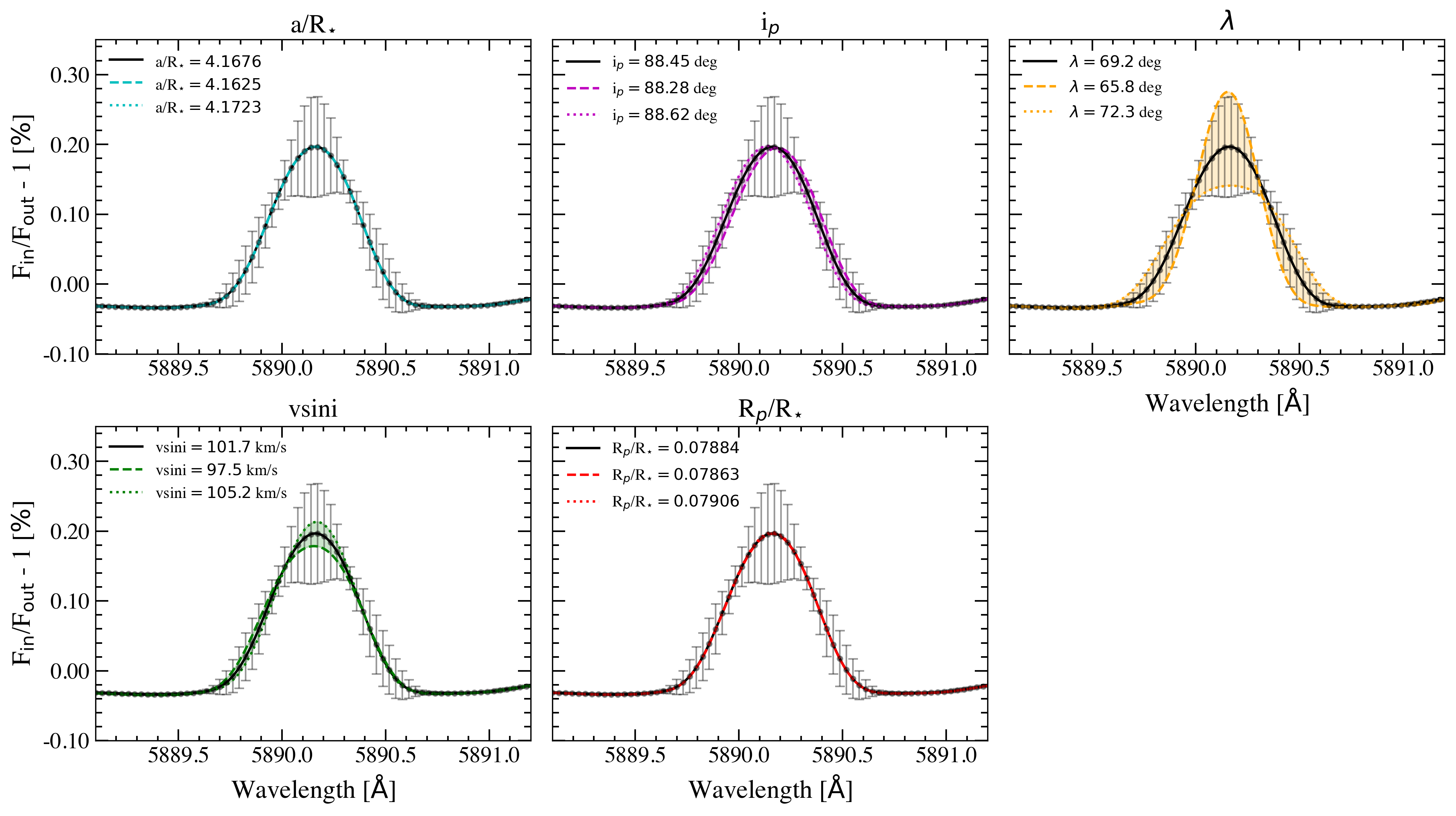}
\caption{Model of the RM and CLV effects in the transmission spectrum of MASCARA-1b around the \ion{Na}{i} D2 line. Each panel shows the model computed with the parameters from \citet{Hooton2021} (black) and the models computed when one of the main parameters is changed to its upper (coloured dotted line) and lower (coloured dashed line) uncertainty value. The error bars of the model derived from the uncertainties of all these parameters are shown in grey.}
\label{fig:RM_uncertain}
\end{figure*}

We explored the possibility that the atmospheric features from the exoplanet are hidden by the RM effect. To do this, we studied the uncertainties of the RM modelling by propagating the error bars of the main system parameters involved in this calculation, that is, $i_p$, $a/R_{\star}$, $\lambda$, $R_p/R_{\star}$, and $v\sin i_{\star}$. In addition, we also analysed the uncertainties induced by the individual parameters in order to find which one contributes more to the overall model uncertainty (see Fig.~\ref{fig:RM_uncertain}). This exercise is performed in a small region around the \ion{Na}{i} D2 line in order to reduce the computational costs. We observe that, in this case, the uncertainties of the model are dominated by the precision of the obliquity measurement ($\lambda = 69.2^{+3.1}_{-3.4}$\,deg; \citealt{Hooton2021}). We note, however, that the uncertainties related to the calculation of the stellar models ($T_{\rm eff}$, $\log g$, [Fe/H], the abundance of each species, etc.) and other line profile asymmetries introduced by effects requiring 3D modelling of the stellar photospheres such as granulation and convective blueshift \citep{CiavassaBrogi2019} are not considered here, and they are expected to significantly increase the model uncertainties.

Assuming that $\lambda$, $i_p$, and $v\sin i_{\star}$ are the dominant sources of uncertainty (see Fig.~\ref{fig:RM_uncertain}), if they are propagated in the RM model calculation around the different single lines analysed in the present work (see Fig.~\ref{fig:TS_res}), we see that the accuracy of the RM model is still not sufficient to robustly disentangle the presence of an absorption in any of the lines. As seen for the \ion{Na}{i} D doublet, the model requires additional sources of uncertainty to be consistent with the observations. However, this is very likely related to the stellar properties such as the abundance of the particular species, among others, as the spectral lines profile of the modelled stellar spectrum do not completely describe the observed MASCARA-1 spectrum. In this same figure, we can also observe that the uncertainties introduced by these parameters are smaller in lines such as \ion{Ca}{ii,} H\&K, and H$\alpha$. This is because these lines are broader and, consequently, the RM effect is broader, too. A change in the $\lambda$ value produces a smaller change of the RM effect in the planet rest frame when compared to the \ion{Na}{i} lines. For those species in the stellar spectrum, a better knowledge of the star and the refinement of the obliquity of the system are needed to robustly detect or refute their presence in the transmission spectrum. Follow-up studies of the RM signature using advanced techniques accounting for the particularities of this star (oblateness and gravity darkening) will be able to precisely measure the obliquity and provide a definite conclusion about the atmosphere of MASCARA-1b.

\subsection{Bulk properties of MASCARA-1b}

MASCARA-1b is one of the densest ultra-hot Jupiters known to date (see Fig.~\ref{fig:ContextMASC1}) with a gravity of $\log g_p = 3.6$ (cgs). This results in a scale height of $258$\,km (assuming $\mu=2.3$ and $T_{eq}=2594$\,K) that corresponds to a transmission (see Fig.~\ref{fig:ContextMASC1}) of only 28\,ppm. Although this precision is reached with ESPRESSO observations for several of the species analysed here (see Table~\ref{tab:upper_lim1}), we find a featureless transmission spectrum.

\begin{figure}[]
\centering
\includegraphics[width=0.48\textwidth]{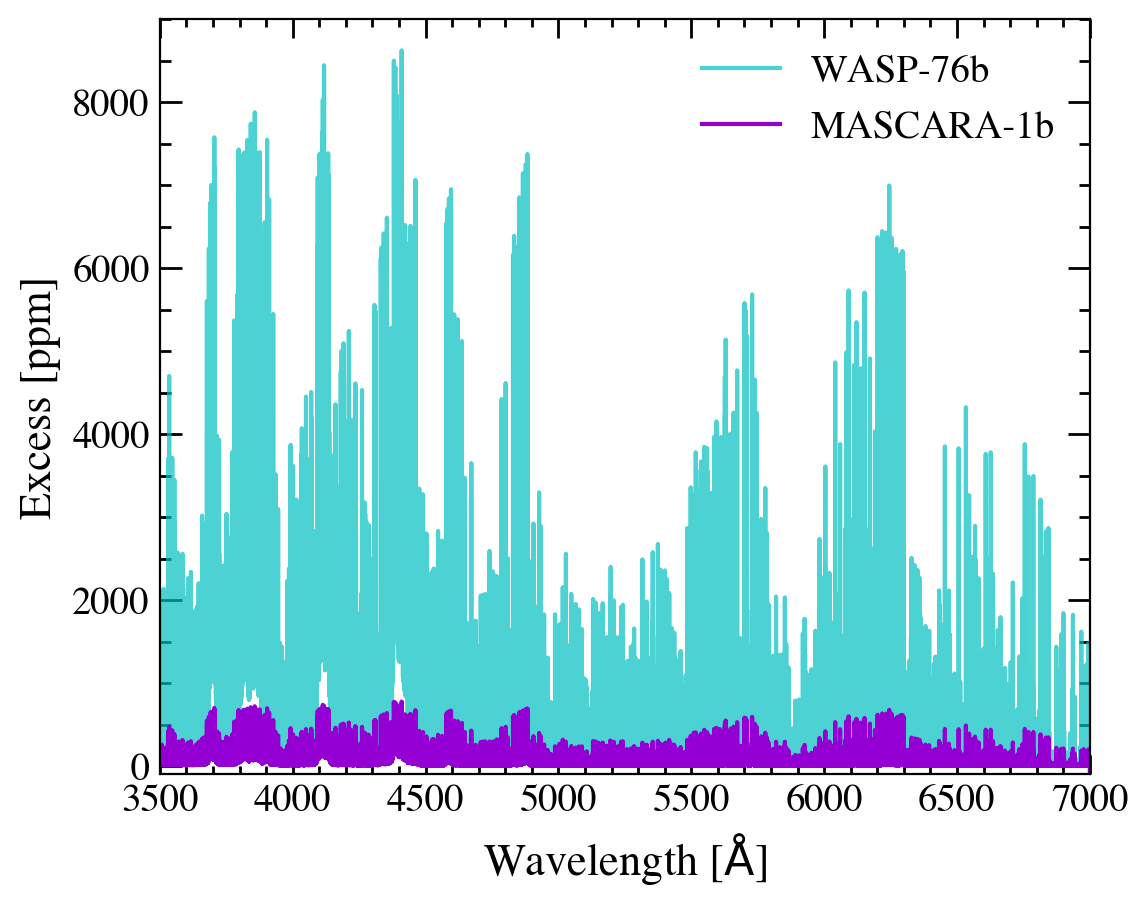}
\caption{Comparison of the V atmospheric model of WASP-76b (cyan) and MASCARA-1b (violet) computed with {\tt petitRADTRANS} assuming their equilibrium temperatures and a volume mixing ratio of $10^{-5}$. In the vertical axis we show $(R_{\lambda}/R_{\star})^2$ with the continuum $(R_p/R_{\star})^2$ subtracted, in units of ppm, for a better comparison of the lines strength. Both models are shown at the ESPRESSO resolution ($\Re=140\,000$) and consider a planet tidally locked rotation of $v_{\rm rot} = 5.3$\,km\,s$^{-1}$ and $v_{\rm rot} = 3.9$\,km\,s$^{-1}$ for WASP-76b and MASCARA-1b, respectively.}
\label{fig:mods_comp}
\end{figure}

We can compare these calculations with those of the well-studied ultra-hot Jupiter WASP-76b, for which several species have been detected using transmission spectroscopy at high spectral resolution (e.g. \citealt{Seidel2019,Tabernero2021b,Ehrenreich2020,Kesseli2021,Kesseli2021b, Casasayas2021b}). Assuming a $T_{\rm eq} = 2228$\,K and $\log g_p = 2.8$ \citep{Ehrenreich2020}, its scale height is $1240$\,km and the transmission is $221$\,ppm. This is an atmosphere about five times more extended than that of MASCARA-1b, and a transmission metric ten times larger. We calculated simulated transmission spectra of V from atmospheric models for these two planets using {\tt petitRADTRANS}, assuming isothermal profiles at their equilibrium temperatures, the same volume mixing ratio of $10^{-5}$, and a cloud deck at 1\,mbar (see Fig.~\ref{fig:mods_comp}). We selected this particular species because it was recently detected in WASP-76b \citep{Kesseli2021b}, and it is not affected by the RM effect in MASCARA-1b. The continuum level of both atmospheric models has been subtracted for a better comparison of the lines' strength. As expected, the strongest \ion{V}{i} lines in WASP-76b simulated transmission spectrum are about ten times more intense than those in MASCARA-1b transmission spectrum.

\section{Conclusions} \label{sec:conclusions}

We report on the analysis of two transit observations of MASCARA-1b with the ESPRESSO spectrograph. In particular, we search for the presence of \ion{Fe}{i}, \ion{Fe}{ii}, \ion{Ca}{i}, \ion{Y}{i}, \ion{V}{i}, \ion{V}{ii}, CaH, and TiO in the atmosphere of MASCARA-1b using the cross-correlation technique, and explore the transmission spectrum around the \ion{Ca}{ii} H\&K, \ion{Na}{i} doublet, \ion{Li}{i}, H$\alpha$, and \ion{K}{i} D1 spectral lines.

Our analysis of the MASCARA-1b transmission spectrum shows a strong impact of the RM effect on all spectral lines present in the stellar spectrum, as observed by \citet{Stangret2021} using HARPS-N observations and non-detections for the remaining species. Due to the system's architecture and the radial-velocity change of the planet, the position of both Doppler shadow and the expected atmospheric absorption of the planet overlap. This scenario can very likely happen in other planetary systems, as also reported for HD~209458b \citep{Casasayas2021}. We discuss the possibility of these species actually being present in the transmission spectrum but hidden by the RM contamination. For this, we compare our results with models of the RM and CLV effects. Unfortunately, using even the most up-to-date and precise ephemerides from CHEOPS \citep{Hooton2021} and the high S/N of ESPRESSO observations, we cannot conclude whether these species are present or not in the transmission spectrum with our analysis due to the differences between the model and the observations. Follow-up studies of the RM signal with more advanced techniques that are able to  account for the effects of this particular star (oblateness and gravity darkening) are needed to reach a definite conclusion about the presence of these species in MASCARA-1b's atmosphere. 
In order to avoid the RM effect, we searched for species not present in the stellar spectrum (\ion{Y}{i}, \ion{V}{i}, \ion{V}{ii}, CaH, and TiO), but none of them show any significant signal. We also discuss the possibility of these non-detections resulting from the low transmission expected for MASCARA-1b due to its high surface gravity. When compared to WASP-76b, for example, the estimated transmission signal for MASCARA-1b is $\sim10$ times smaller. Nevertheless, the ESPRESSO observations have a high enough S/N that we should have detected some of the species.

Although MASCARA-1b is not optimal for transmission spectroscopy studies at high spectral resolution, due to its high density and the impact of the Doppler shadow, it is probably a good target for emission spectroscopy studies due to its high day-side temperature ($T_{\rm day} =3062$\,K; \citealt{Hooton2021}) and the brightness of its host star (V=8.3\,mag). Emission spectroscopy techniques (e.g. \citealt{Yan2020, Pino2020, Nugroho2020FeIe, Cont2021}) open a door to atmospheric studies of these particularly complicated planets. In addition, other studies such as the chromatic Doppler tomography technique (\citealt{Borsa2016,Santos2020,EsparzaBorges2021}) may also help provide insights into their atmospheric composition.

\begin{acknowledgements}

We acknowledge funding from the European Research Council under the European Union’s Horizon 2020 research and innovation programme under grant agreement No. 694513. This work made use of PyAstronomy \citep{PyAstronomy2019ascl.soft06010C} and of the VALD database, operated at Uppsala University, the Institute of Astronomy RAS in Moscow, and the University of Vienna. The INAF authors acknowledge financial support of the Italian Ministry of Education, University, and Research with PRIN 201278X4FL and the "Progetti Premiali" funding scheme. JIGH, RR, CAP and ASM acknowledge financial support from the Spanish Ministry of Science and Innovation (MICINN) project PID2020-117493GB-I00. ASM, JIGH and RR also acknowledge financial support from the Government of the Canary Islands project ProID2020010129. ASM acknowledges financial support from the Spanish MICINN under 2018 Juan de la Cierva programme IJC2018-035229-I. This work was supported by Fundação para a Ciência e a Tecnologia (FCT) and Fundo Europeu de Desenvolvimento Regional (FEDER) via COMPETE2020 through the research grants UID/FIS/04434/2019; UIDB/04434/2020; UIDP/04434/2020; PTDC/FIS-AST/32113/2017 \& POCI-01-0145-FEDER-032113; PTDC/FIS-AST/28953/2017 \& POCI-01-0145-FEDER-028953; PTDC/FIS-AST/28987/2017 \& POCI-01-0145-FEDER-028987. O.D.S.D. is supported in the form of work contract (DL 57/2016/CP1364/CT0004) funded by FCT. E. E-B. acknowledges financial support from the European Union and the State Agency of Investigation of the Spanish Ministry of Science and Innovation (MICINN) under the grant PRE2020-093107 of the Pre-Doc Programme for the Training of Doctors (FPI-SO) through FSE funds. R. A. is a Trottier Postdoctoral Fellow and acknowledges support from the Trottier Family Foundation. This work was supported in part through a grant from FRQNT. This work has been carried out in the frame of the National Centre for Competence in Research PlanetS supported by the Swiss National Science Foundation (SNSF). The authors acknowledge the financial support of the SNSF. This project has received funding from the European Research Council (ERC) under the European Union's Horizon 2020 research and innovation programme (project {\sc Spice Dune}, grant agreement No 947634). FB acknowledges support from PRIN INAF 2019. V.A. acknowledges the support from FCT through Investigador FCT contract nr.  IF/00650/2015/CP1273/CT0001. M.L. acknowledges support of the Swiss National Science Foundation under grant number PCEFP2\_194576.

\end{acknowledgements}

\bibliographystyle{aa} 
\bibliography{biblio} 

\onecolumn
\begin{appendix} 

\section{Additional figures} \label{app:add_fig}


\begin{figure*}[h]
\centering
\includegraphics[width=1\textwidth]{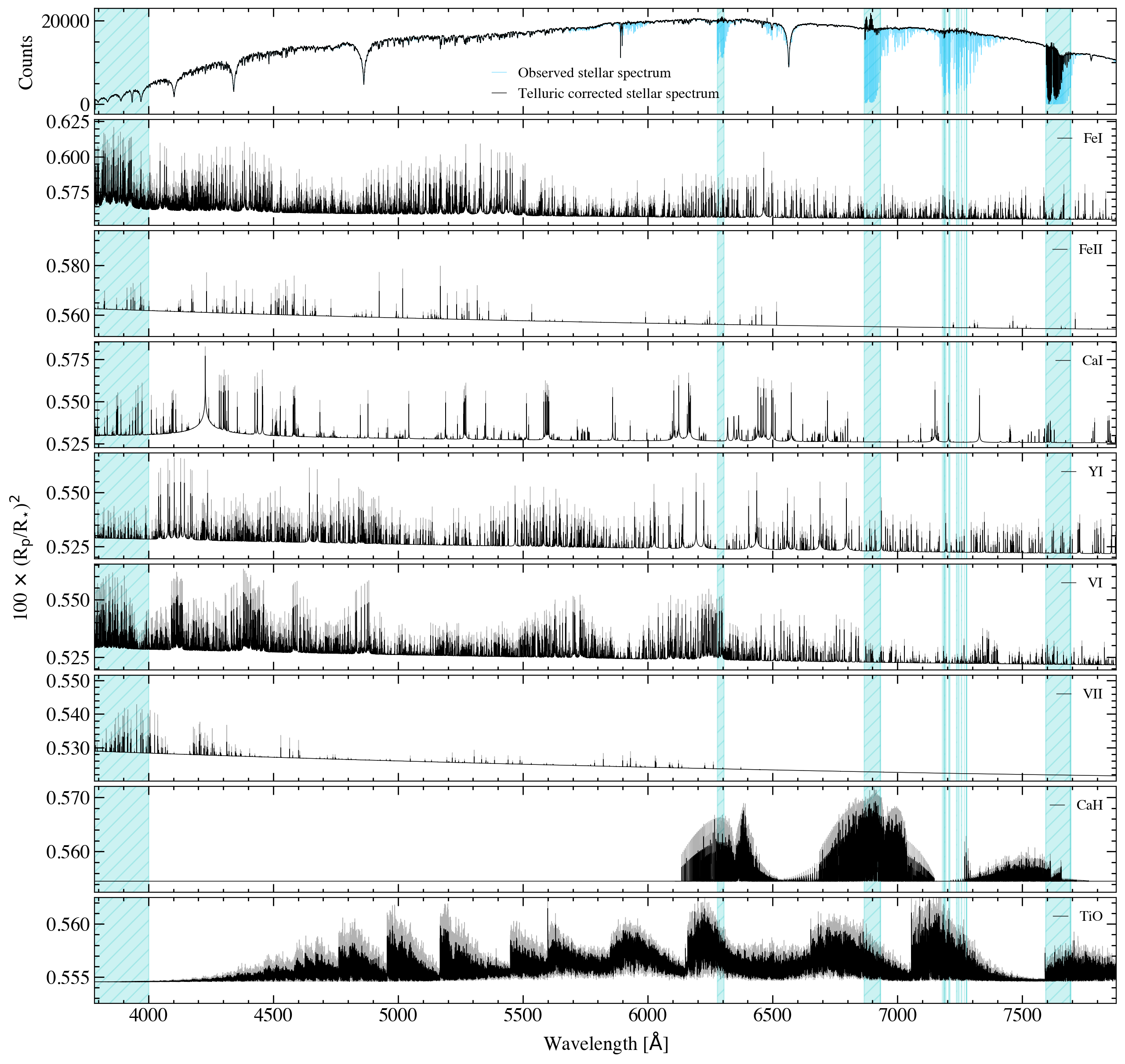}
\caption{Atmospheric models of MASCARA-1b computed with {\tt petitRADTRANS} with (black) and without (grey) considering the broadening produced by the tidally locked rotation and the movement of the planet in one exposure. In the top row we show the mean observed stellar spectrum (blue) and the telluric corrected spectrum (black) as reference. The light blue shadowed regions indicate those wavelengths that are not considered in the cross-correlation analysis due to telluric residuals or a low S/N.}
\label{fig:CC_mods}
\end{figure*}

\newpage

\begin{figure*}[]
\centering
\includegraphics[width=0.9\textwidth]{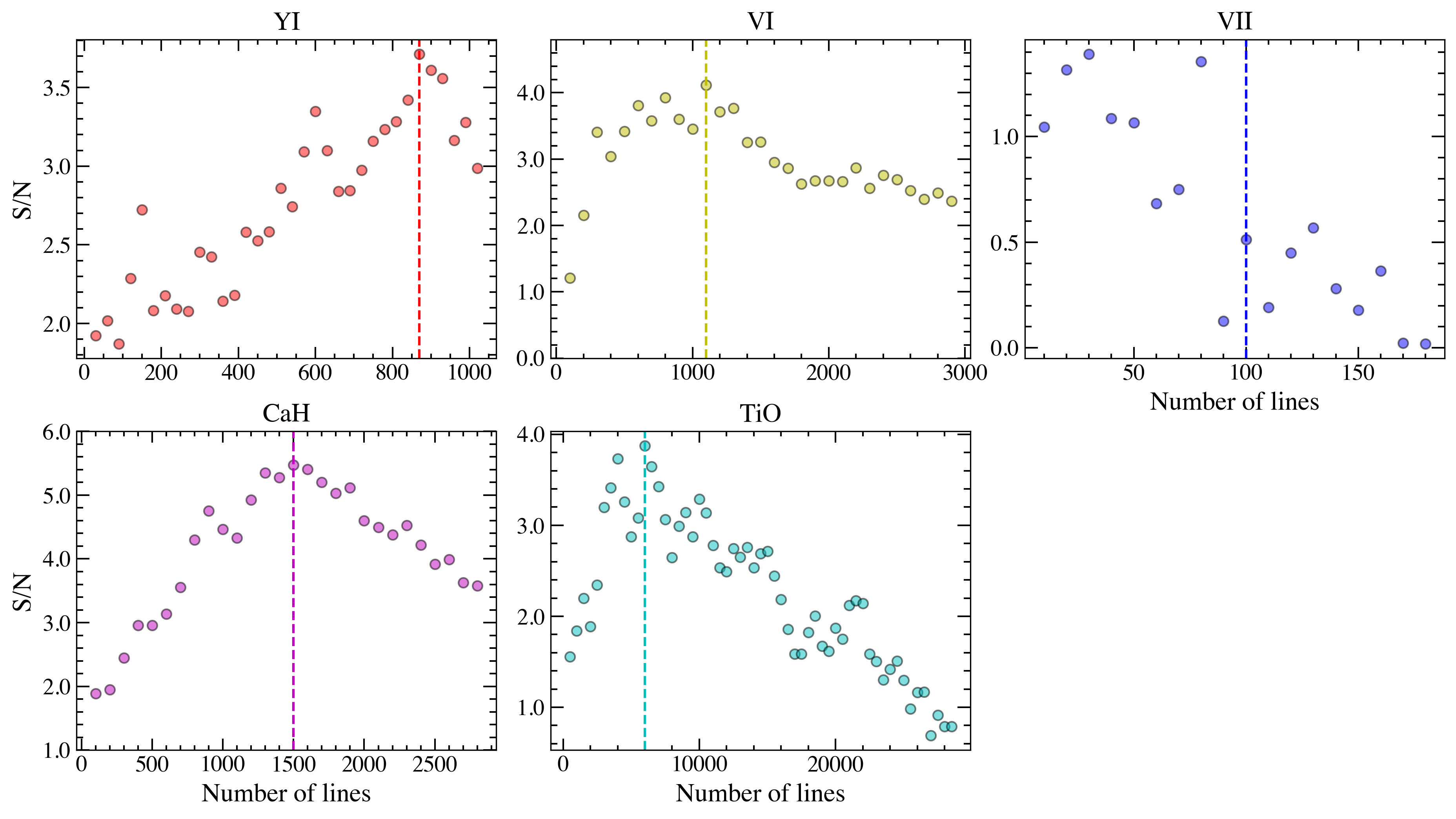}
\caption{Evolution of the significance of the recovered signal (injected at +135\,km\,s$^{-1}$) when increasing the number of lines included in the binary mask, ordered in terms of strength. The vertical dashed lines indicate the number of lines selected to estimate the upper limits. These results are obtained as the combinations of the two transit observations. If the signal is not recovered, we assume a minimum of 100 lines (e.g. \ion{V}{ii}).}
\label{fig:Nlines}
\end{figure*}

\begin{figure*}[]
\centering
\includegraphics[width=0.48\textwidth]{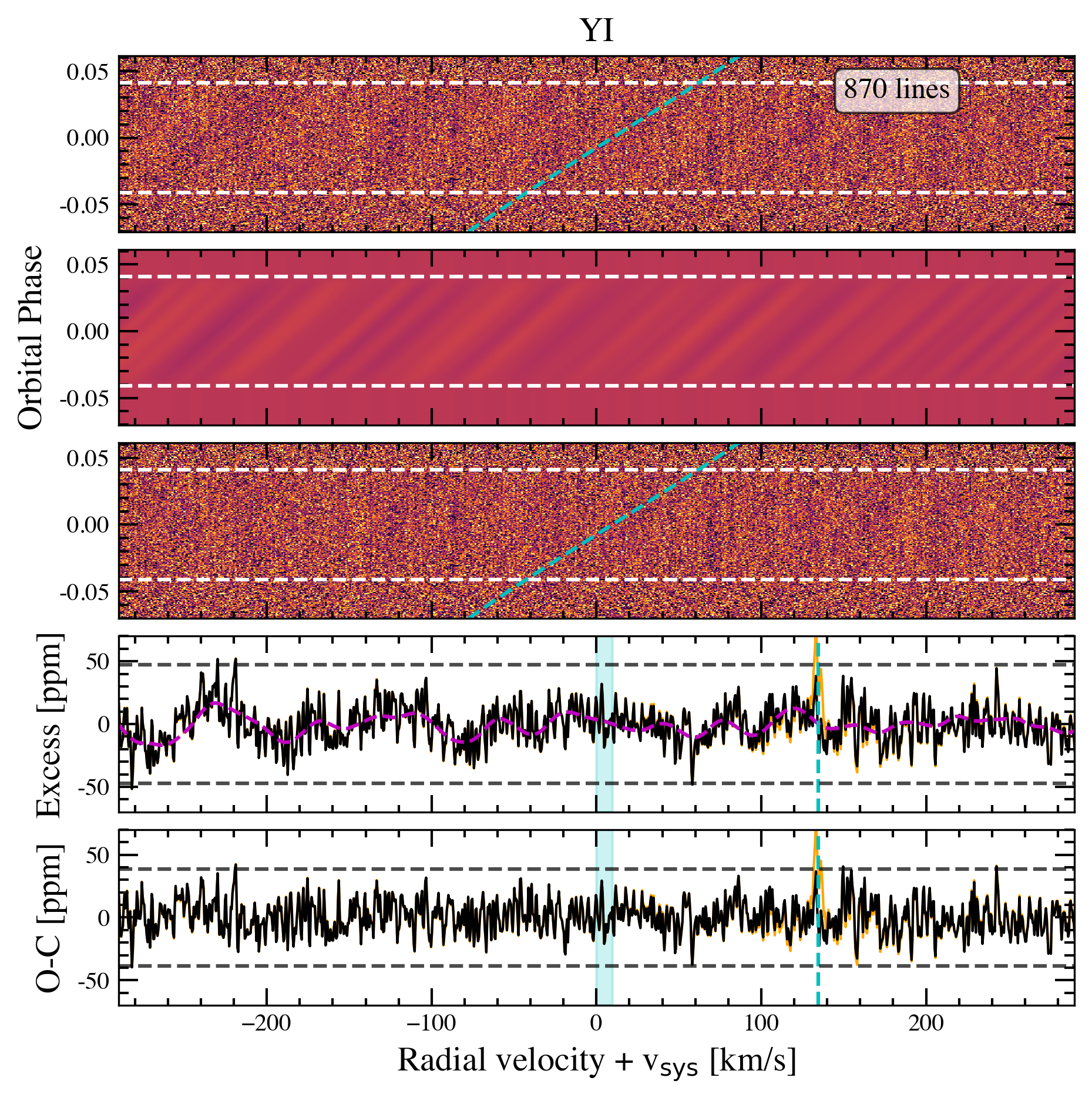}
\includegraphics[width=0.48\textwidth]{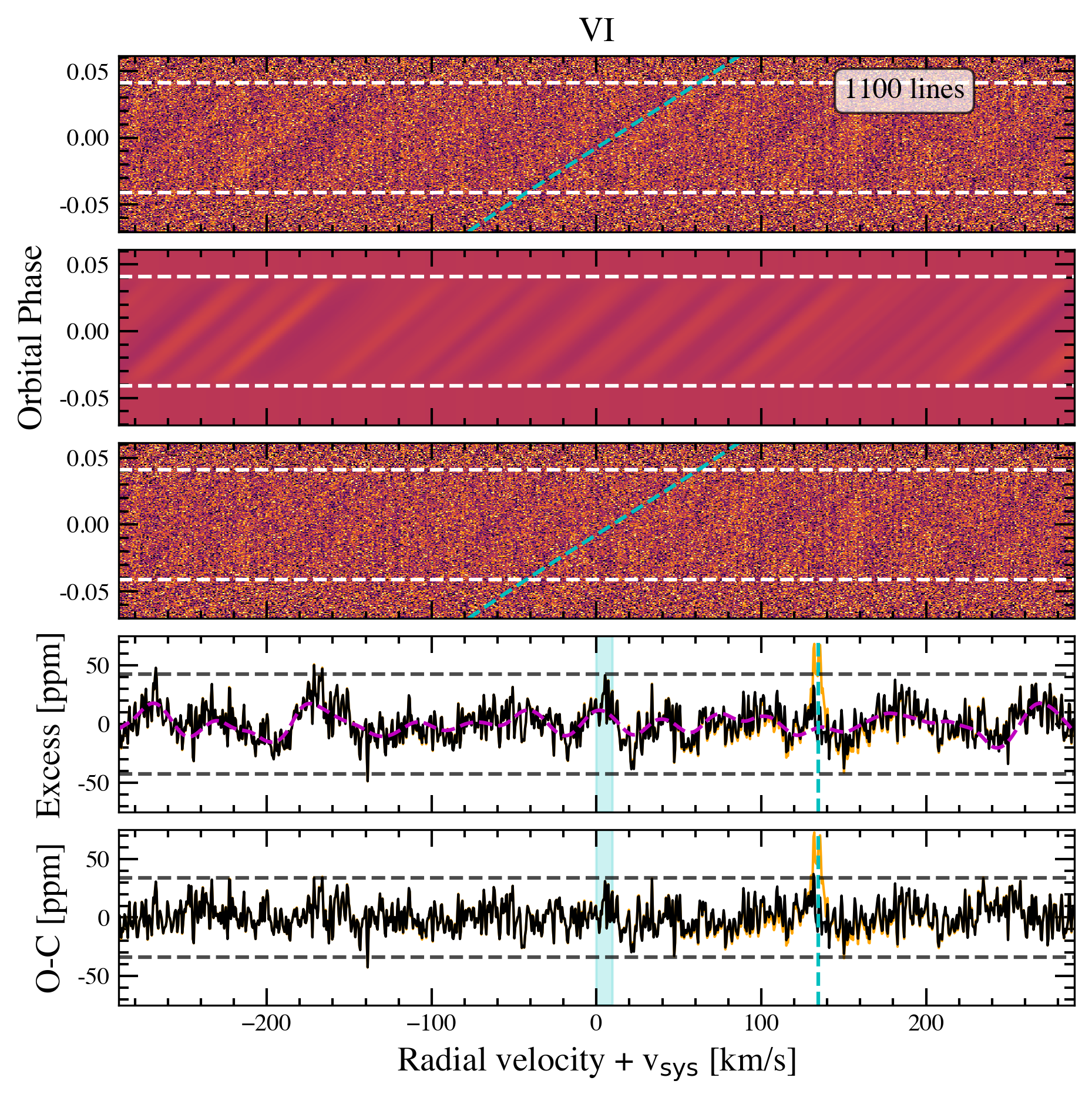}
\includegraphics[width=0.48\textwidth]{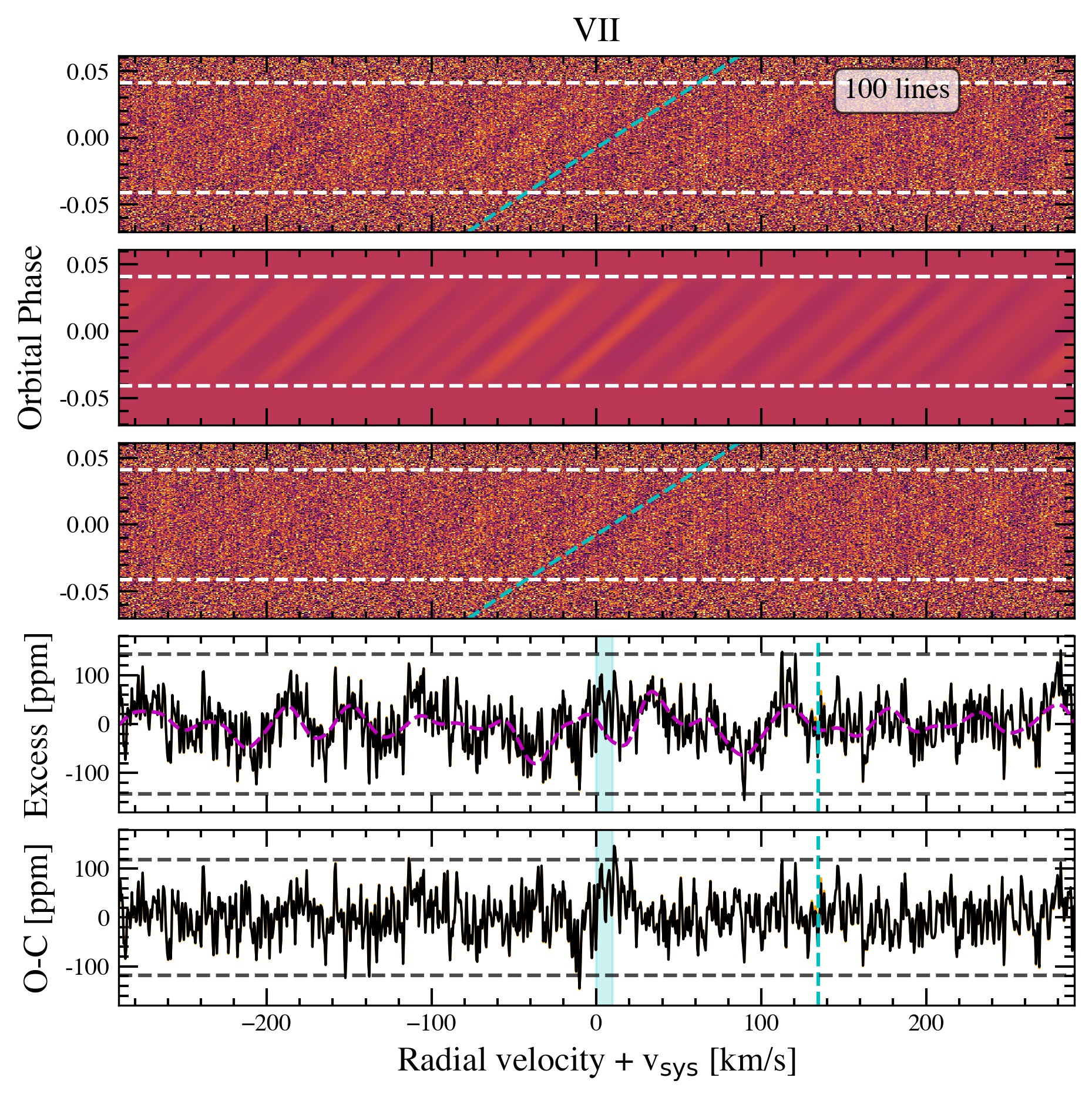}
\includegraphics[width=0.48\textwidth]{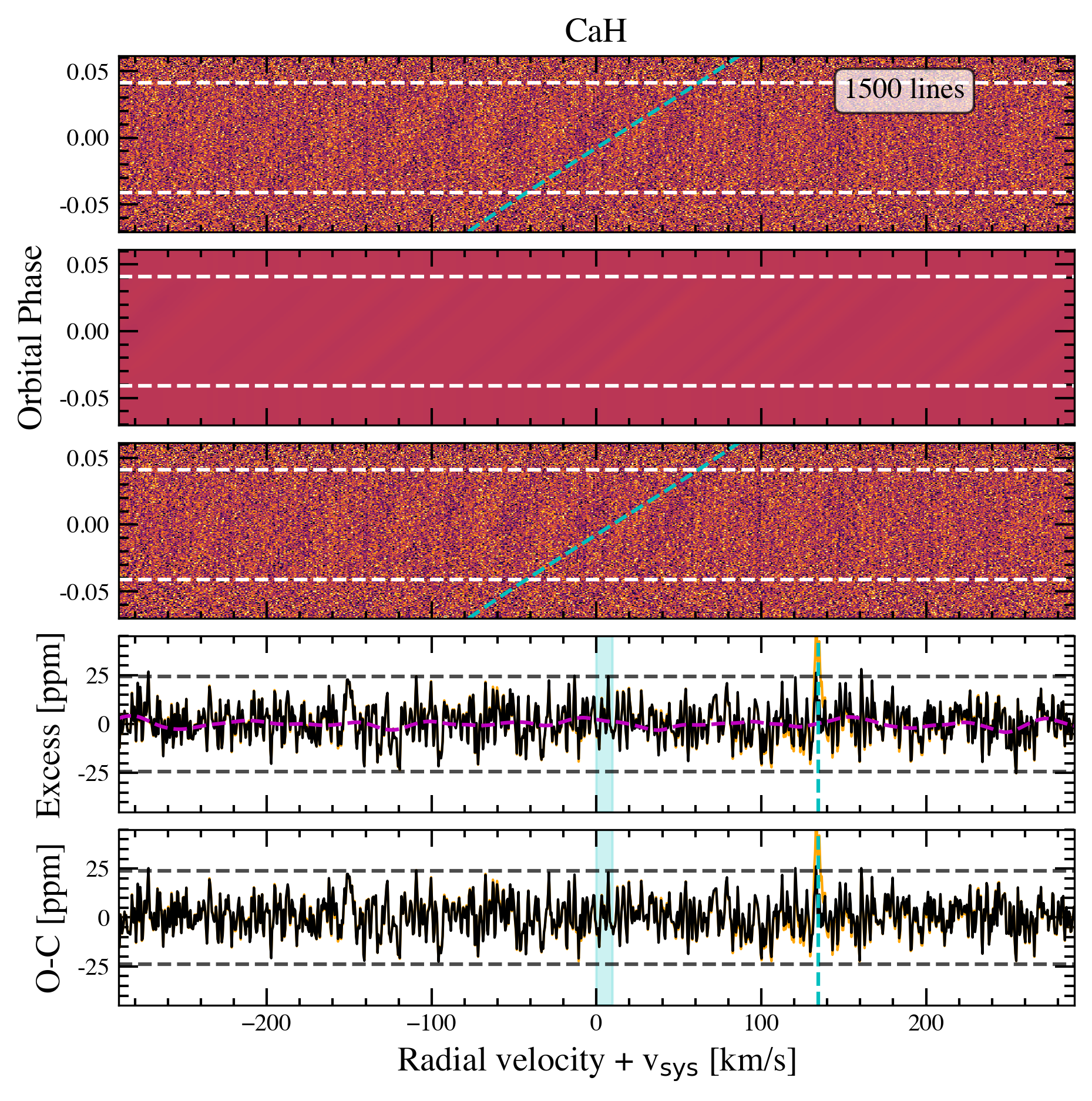}
\caption{Cross-correlation results of \ion{Y}{i}, \ion{V}{i}, \ion{V}{ii}, and CaH obtained with a binary mask including a particular number of strong spectral lines (see Fig.~\ref{fig:Nlines}). Each panel corresponds to one species, and the number of lines used in the binary mask is indicated in the top right corner of each panel. {\em First row (top)}: Cross-correlation maps. {\em Second row}: Cross-correlation maps obtained with the modelled spectra containing the RM and CLV effects on the stellar lines. {\em Third row}: Cross-correlation maps after subtracting the models (second row) to the observations (first row). The white horizontal lines indicate the first and last contacts of the transit. The dashed cyan lines indicate the planet radial-velocity change during the observations assuming $v_{\rm sys}= 9.3$\,km\,s$^{-1}$. {\em Forth row}: Combination of the in-transit results in the planet rest frame before the model subtraction (first row). The violet line shows the expected RM and CLV effect deformation of the lines in this region. {\em Fifth row}: Combination of the in-transit results in the planet rest frame after the model subtraction (third row). The $\pm3\sigma$ noise level is shown in horizontal dashed black lines. The orange line is the cross-correlation result including the injected atmospheric signal at $+135$\,km\,s$^{-1}$, whose position is indicated with a vertical dashed cyan line. The light blue region shows the $v_{\rm sys}$ range where the signal from exoplanet atmosphere is expected.}
\label{fig:RM_corr1}
\end{figure*}

\begin{figure*}[]
\centering
\includegraphics[width=0.47\textwidth]{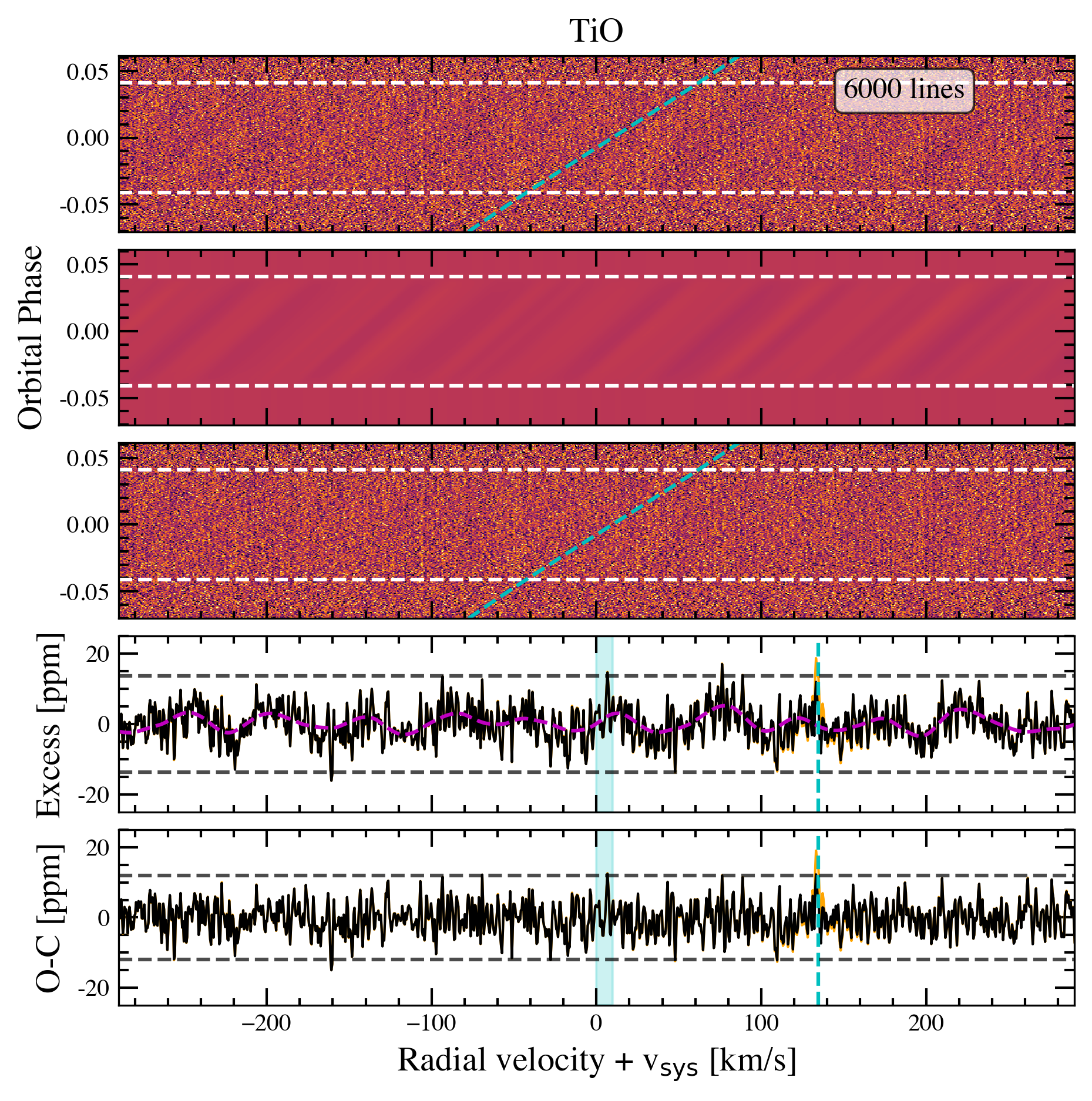}
\caption{Same as Fig.~\ref{fig:RM_corr1} but for TiO.}
\label{fig:RM_corr2}
\end{figure*}

\end{appendix}

\end{document}